\documentclass[acmsmall,noacm,authorversion]{acmart}

\setcopyright{none} 
\renewcommand\footnotetextcopyrightpermission[1]{}

\usepackage{multirow}
\usepackage{wrapfig,booktabs}
\usepackage{colortbl}
\usepackage{subcaption} 
\usepackage{amsmath}
\usepackage[ruled,linesnumbered]{algorithm2e}
\AtBeginDocument{%
  }

\newcommand{\ie}{{\em i.e., }}
\newcommand{\eg}{{\em e.g., }}

\begin{document}

\title{\textit{FastFlow}: Early Yet Robust Network Flow Classification using the Minimal Number of Time-Series Packets}\thanks{This is the preprint version of our paper \cite{RBabariaSIGMETRICS25} accepted at ACM SIGMETRICS 2025.}

\author{Rushi Jayeshkumar Babaria}
\affiliation{%
  \institution{University of New South Wales}
  \city{Sydney}
  \state{NSW}
  \country{Australia}
}
\author{Minzhao Lyu}
\affiliation{%
  \institution{University of New South Wales}
  \city{Sydney}
  \state{NSW}
  \country{Australia}
}\thanks{Corresponding author: Minzhao Lyu (minzhao.lyu@unsw.edu.au)}

\author{Gustavo Batista}
\affiliation{%
  \institution{University of New South Wales}
  \city{Sydney}
  \state{NSW}
  \country{Australia}
}
\author{Vijay Sivaraman}
\affiliation{%
  \institution{University of New South Wales}
  \city{Sydney}
  \state{NSW}
  \country{Australia}
}

\begin{abstract}
Network traffic classification is of great importance for network operators in their daily routines, such as analyzing the usage patterns of multimedia applications and optimizing network configurations. Internet service providers (ISPs) that operate high-speed links expect network flow classifiers to accurately classify flows early, using the minimal number of necessary initial packets per flow. These classifiers must also be robust to packet sequence disorders (drops and retransmissions) in candidate flows and capable of detecting unseen flow types that are not within the existing classification scope, which are not well achieved by existing methods.
In this paper, we develop \textit{FastFlow}, a time-series flow classification method that accurately classifies network flows as one of the known types or the unknown type, which dynamically selects the minimal number of packets to balance accuracy and efficiency. Toward the objectives, we first develop a flow representation process that converts packet streams at both per-packet and per-slot granularity for precise packet statistics with robustness to packet sequence disorders. Second, we develop a sequential decision-based classification model that leverages LSTM architecture trained with reinforcement learning. Our model makes dynamic decisions on the minimal number of time-series data points per flow for the confident classification as one of the known flow types or an unknown one. We evaluated our method on public datasets and demonstrated its superior performance in early and accurate flow classification. Deployment insights on the classification of over 22.9 million flows across seven application types and 33 content providers in a campus network over one week are discussed, showing that \textit{FastFlow} requires an average of only 8.37 packets and 0.5 seconds to classify the application type of a flow with over 91\% accuracy and over 96\% accuracy for the content providers.
\end{abstract}



\maketitle

\section{Introduction}
Network operators that offer Internet access services to broadband and mobile clients, or manage network infrastructure for large enterprises and campuses, are eager to gain visibility into the bandwidth demand by different applications, such as video streaming, online gaming, and conferencing, and their service providers (\eg YouTube and Netflix for video streaming), along with the associated experience received by the end-users, for the following major business and operational reasons.
First, knowing which network services and providers are becoming popular in their deployment geography can directly impact the business planning of a network operator, such as creating selective bundles and premium subscription plans that have the potential to increase its profit margin. Second, having visibility into the experience associated with each application session facilitates proactive network troubleshooting to increase customer satisfaction and reduce potential revenue loss. Third, providing contextual labels for network flows through a network can serve as signals for dynamic performance optimization techniques, such as prioritizing network flows of latency-sensitive applications like video conferencing via low-latency 5G slices, and using network APIs for video streaming flows to guarantee sufficient bandwidth allocations.

Internet service providers (ISPs) deploy classifiers that categorize network flows into their application types for usage accounting and subsequent user experience measurement \cite{zoom_IMC_passive_2022,zoom_webex_meet_measurement,Minzhao_IMC_User_platform,Zoom_PAM_network_level_view,IMC_measurement_network_util_conf}. For practical deployment at large networks that can serve hundreds of thousands of users using diversified applications supported by millions of concurrent network flows, flow classifiers are expected to not only provide accurate flow labels that are of interest to network operators but also produce classification results for each flow at the earliest possible time, ideally by just seeing the initially arrived packets of a flow. This approach spares computing resources to classify more flows in the queue and quickly enables subsequent monitoring tasks (\eg inference on application user experience \cite{zoom_webex_meet_measurement,zoom_IMC_passive_2022}) that take the flow classification results as a prerequisite.

Network traffic classifiers have started shifting from rule-based mechanisms that match flow metadata, such as port numbers and server name indication (SNI) fields in TLS handshake payloads, which have become less effective due to increasingly encrypted application traffic and the use of randomized or non-standardized port numbers. Instead, statistical models are now being used to classify network flows based on their time-series volumetric profiles, which are determined by the actual content carried per application and are agnostic to metadata obfuscation and encryption. However, for a large network like an ISP that can have millions of concurrent flows to be quickly classified for downstream tasks such as usage accounting and performance measurement in real-time, legacy metadata-based approaches can classify a candidate flow by its first or certain signaling packet(s), while time-series statistical models often require a lengthy time-series input (\eg packets) before concluding a confident flow classification result.

Therefore, to enable the practical deployment of time-series statistical flow classifiers in large-scale networks, prior research works \cite{zoom_IMC_passive_2022,on_early_traffic_classification, early_classification_application_negotiation,effective_packets, lstm_10_20_30, lyu2023metavradar, knowThyLag, minzhao_cloud_gaming, ggfast} have developed such models using a fixed number of initial packets in a flow that carry static signatures in initialization requests pertinent to a certain application (\eg video streaming or conferencing) or content provider. However, given the diversified flow types in an ISP network, each with its static initial content carried by a different number of initial packets, having this number fixed for every flow cannot always produce reliable classification. Specifically, with a fixed number, unpredictable packets carrying dynamic user content may be included for flows that inherently require a smaller number of initial packets for classification, and initial packets carrying static flow content may not be sufficiently captured for flows that require a larger number. Moreover, packet drops and retransmissions, which are common in network communications, further render a fixed number of initial packets for classifying diversified flows in realistic network environments ineffective.

To address this gap in network flow classification for large-scale networks, a run-time optimization is required to determine the minimal number of packets to balance prediction accuracy and system efficiency \cite{CATO}. An insufficient number of time-series inputs can cause inaccurate results, whereas a number more than the just enough value can lead to delays in classification. Such optimization objectives are formulated as a sequential decision-making problem by the machine learning community and approached by reinforcement learning techniques \cite{rl_in_robotics,go_using_rl}. Those provide inspiration for us to develop a time-series network flow classifier with sequential decision-making capability trained by reinforcement learning, which can determine the minimal number of time-series data points at run-time to confidently classify each candidate flow.

In this paper, we develop \textit{\textbf{FastFlow}}, a time-series flow classification method to accurately classify network flows for their application types and content providers using the estimated minimal number of packets carrying initial static content pertinent to their types. The minimal number is dynamically estimated for each candidate flow at run-time. As will be overviewed in \S\ref{sec:FastFlowOverview}, our method \textit{\textbf{FastFlow}} represents raw packet streams of a flow as well-curated data sequences at dual granularity of packets and slots (\ie time intervals), and uses purposely designed time-series classifiers leveraging long-short-term memory (LSTM) architecture tuned by reinforcement learning techniques for run-time estimation on the minimal number of time-series inputs for classifying each flow. The method is inherently capable of detecting unknown flows as outliers instead of mislabeling them as one of the known types, and is robust to packet sequence disorders within a flow due to packet drops and retransmissions, which are common cases during deployment at large networks. Our contributions in this paper are three-fold.

Our \textbf{first contribution} (\S\ref{sec:Representation}) defines a \textbf{dual-grained time-series flow data sequence} to represent volumetric statistics of a network flow as formal inputs to our time-series flow classifiers. The fine-grained flow data sequence is curated for precise packet statistics of a candidate flow covering not only packet sizes but also contextual information including directions and inter-arrival times which serve as strong indicators for flow types when the packet sequence perfectly matches its expected norm. It is complimented by a coarse-grained time-series data sequence that consists of volumetric statistics of a flow per slot, which exhibits better statistical robustness to packet sequence disorders caused by packet drops and retransmissions through interval-based aggregation. The joint use of both types of time-series data sequence via a run-time selection process provides quality representation of a flow with both precision and robustness to sequence disorder of its arriving packets.

In the \textbf{second contribution} (\S\ref{sec:TimeSeriesModel}), we develop a \textbf{time-series flow classifier architecture} that leverages long-short term memory cells (LSTM) to classify network flows on the dual-grained time-series flow data sequence describing statistics of packets arrived at runtime. The classifier is trained by reinforcement learning techniques as a sequential decision making model with inference functions (\ie linear layers) flexible on prediction timestamps and an intermediate `unknown' flow type. The trained classifiers can dynamically estimate the minimal number of time-series data points (\eg packets) to confidently classify each candidate flow. To enhance the accuracy of our trained classifiers in detecting unknown flow types that have their statistical characteristics deviating from the known types, we develop a training data augmentation technique that compensates the commonly absent unknown flow characteristics to improve the convergence of statistical boundaries in the trained models for classifying known flow types.

The \textbf{third contribution} (\S\ref{sec:Evaluation}) evaluates the classification performance of \textit{FastFlow}. We start by comparing \textit{FastFlow} with \textbf{state-of-the-arts methods and ablation alternatives} on three popular public network flow datasets from university and ISP labs, showcasing the performance of \textit{FastFlow} in both simple (over 98\% accuracy using on average 4 initial packets in 0.03 seconds per flow) and complex (over 86\% accuracy using on average 13 initial packets in 0.6 seconds per flow) classification tasks. We then deploy \textit{FastFlow} in a large campus network. Using flow labels from a commercial network traffic classification system as estimated ground truths, we demonstrate \textbf{deployment insights in the realistic network environment} including flow classification performance of seven popular application types and 33 content providers. On average, only 8.37 initial packets of a flow are needed to classify its application type with an accuracy higher than 91\%. Among application types, conferencing flows are detected with the highest accuracy of 99.82\% using 3 to 17 initial packets in approximately 1.34 seconds. \textit{FastFlow} also achieves superior performance in classifying content providers of each application type, an accuracy of 97\% is achieved to classify video streaming providers using the first 3 to 6 packets of each flow.

\section{Related Work}\label{sec:RelatedWorks}
Classifying encrypted network flows using time-series data for application classification \cite{stastical_method_yet_another, mobile_app_deep_learning} and network anomaly detection \cite{attack_detection, ml_classifiers_intrusion_detections} has been a popular topic given its importance for network operation. Despite the large number of prior works that develop or fine-tune classification models \cite{survey_paper_1, survey_paper_2,survey_paper_3}, many existing works focus on cost-effectively representing time-series flow data and developing methods for early classification of network flows \ie using the least number of time-series data points for accurate prediction, which are closely related to our work.

\textbf{Time-Series Flow Data Representation:} To represent a network flow that consists of a series of upstream and downstream packets arrived at different timestamps, some prior works \cite{time_related_vpn_traffic,CNN_LSTM_combo_fast,Stacked_LSTM,stastical_clustering_protocols_ood_as_well,Gaussian_model_for_fast_classification,prototypical_networks_paper,robust_network_traffic_classification,time_related_clustering,clustering_based_early_classification,unsupervised_cyber_detection,time_related_machine_learning,stastical_method_yet_another,time_related_SVM,old_statical_classification_1,time_related_machine_learning,stastical_clustering_protocols_ood_as_well} use aggregated metrics such as packet inter-arrival time and packet length for a group of packets (such as by a fixed time interval or for a certain number of packets), which lose fine-grained packet-level statistics for accurate classification. To address this issue, other research works \cite{GRU_feature_mobilenet,n_print,flow_pic,GAN_for_low_samples,RNN_CNN_classification,1D_CNN,CNN_LSTM_2} choose to preserve a sequence at per-packet granularity by capturing inter-arrival time and packet size of each packet without aggregation. The classification models trained on such precise sequences can perform poorly in operational networks as packets in a flow do not always come in precise sequence due to packet drops and retransmission. This problem is especially severe for works \cite{effective_packets,early_classification_application_negotiation,ggfast} that classify a flow by its first few packets.
In our work, we propose a flow representation process that collectively uses per-packet representation for accurate inference in general cases and time-series metrics aggregated per slot (\ie time interval) that provide resilience to candidate flows with disordered packet sequence due to packet drops and retransmission.

\textbf{Time-Series Early Classification of Network Flow:} A cost-effective classification of network flows in high-speed telecommunications networks often requires a decision to be taken after inspecting a small number of packets. To balance flow classification accuracy and cost, prior works use a unified number of packets or time intervals for all flows in their classification methodologies \cite{ggfast,flow_pic,Gaussian_model_for_fast_classification,effective_packets,prototypical_networks_paper,lstm_10_20_30}. For example, \textit{GGFast} \cite{ggfast} uses the first 50 packets to classify each flow and \textit{Flowpic} \cite{flow_pic} specifies a constant 10-second time interval limit. In addition, the works in \cite{Gaussian_model_for_fast_classification,on_early_traffic_classification,early_classification_application_negotiation,timely_classification_on_fly,RPC_compare_11_15,sliding_window_N_packets} evaluate the performance of their flow classifiers with various numbers of packets or time intervals. Notably, in \cite{hardEasy}, the authors classify flows using their first several initial packets. However, this number of packets is a hyper parameter optimized on each training dataset, \ie deployment network, rather than on a per flow level.
While such fine-tuned static number of inputs for a certain classification objective (\eg streaming video providers) in one deployment environment can perform well, such methods can be difficult to generalize for changing needs of flow classification objectives (\eg including new application types such as metaverse VR applications \cite{lyu2023metavradar} or new content providers). Such variations in flow classification objectives are common in network operation, thus, having a static number of data points for all flows may not always provide optimal performance. In addition, the number of input data required to classify each flow can also be impacted by dropped or retransmitted packets in a practical environment.
To address the limitation, our flow classification dynamically estimates the minimal number of input packets or time intervals for each candidate flow by equipping the time-series classifiers with a sequential decision-making capability trained by reinforcement learning, which is the first of such attempts (to the best of our knowledge) in network traffic classification.

\section{Overview of the \textit{FastFlow} Time-Series Flow Classification Method}\label{sec:FastFlowOverview}

This section discusses key requirements for flow classifications in large networks that are not fully addressed by state-of-the-arts (\S\ref{sec:flowClassificationRequirements}), which motivate our \textit{FastFlow} method overviewed in \S\ref{sec:method}.

\subsection{Key Requirements for Time-Series Flow Classification in Large Networks}\label{sec:flowClassificationRequirements}

Network operators expect that their flow classifiers have decent prediction performance and low system costs \cite{CATO}. Specifically, three major requirements can determine the practicality of flow classification methods deployed in large networks with massive throughput scaling to millions of concurrent flows: (i) early classification with the estimated minimal length of time-series packet statistics, (ii) detecting unknown flow types, and (iii) robustness to packet sequence disorders. Table~\ref{tab:comparison} provides a comparison of state-of-the-art methods with respect to these three requirements.

First, operators seek to reduce computational resource consumption in processing packets that can be amplified in magnitude when flow classifiers are deployed for millions of concurrent flows in a large network. They also expect each flow to be quickly classified for precise post-prediction telemetry. Therefore, early classification that aims to provide a prediction for each flow as early as possible while maintaining classification accuracy is necessitated. This requires the estimation of an \textbf{minimal number of time-series data points} (\eg packets or slot statistics) for classifying each candidate flow. An unnecessarily large input size may reduce classification speed, whereas an insufficiently small number can lead to inaccurate classification results. State-of-the-art methods either use a fixed input size, such as first 10 packets \cite{Gaussian_model_for_fast_classification}, a unified input size for each deployment network (\eg \cite{ggfast}) or per application type regardless of the possible variations and complexities in flow profiles during practical deployment.

\begin{table}[t!] 
    \centering
        \fontsize{8}{10}\selectfont
        \caption{Qualitative comparison between \textit{FastFlow} and the state-of-the-art in terms of the three key requirements of time-series flow classification for deployment in large networks.}
    \begin{tabular}{|l|l|c|c|}
        \hline
 
        \rowcolor[rgb]{ .906,  .902,  .902}	\textbf{Method}  &\textbf{Number of time-series data points} & \textbf{Unknown flow type} & \textbf{Packet sequence disorder}  \\ 
        \hline
        \cite{ggfast} & \cellcolor{pink} a fixed input size in the method  & \cellcolor{lime} able to detect & \cellcolor{pink}  \textbf{not} robust \\
        \hline
        \cite{lessons_learned_from_commercial} & \cellcolor{yellow} a unified input size \textbf{per TCP/UDP}  & \cellcolor{lime} able to detect & \cellcolor{pink}  \textbf{not} robust \\
        \hline
        \cite{n_print}  & \cellcolor{pink} a fixed input size in the method &\cellcolor{pink}  \textbf{not} able to detect & \cellcolor{pink} \textbf{not} robust \\
        \hline
        \cite{robust_network_traffic_classification} & \cellcolor{yellow} a unified size \textbf{per network} & \cellcolor{lime} able to detect & \cellcolor{pink} \textbf{not} robust \\
        \hline
        \cite{prototypical_networks_paper} & \cellcolor{pink} a fixed input size in the method & \cellcolor{lime} able to detect &\cellcolor{lime} robust\\
        \hline
        \cite{Gaussian_model_for_fast_classification} & \cellcolor{yellow} a unified size \textbf{per network}& \cellcolor{pink} \textbf{not} able to detect &\cellcolor{pink}  \textbf{not} robust \\
        \hline
        \cite{early_classification_application_negotiation} & \cellcolor{yellow} a unified size \textbf{per application type}& \cellcolor{pink} \textbf{not} able to detect & \cellcolor{pink} \textbf{not} robust \\
        \hline
        \cite{effective_packets} & \cellcolor{yellow} a unified size \textbf{per network}& \cellcolor{pink} \textbf{not} able to detect & \cellcolor{pink} \textbf{not} robust \\
        \hline
        \cite{unknown_flow_detection_clustering} & \cellcolor{yellow} a unified size \textbf{per network}& \cellcolor{lime} able to detect & \cellcolor{lime}robust \\
        \hline
        \textit{\textbf{FastFlow}} &\cellcolor{lime} an estimated minimal size \textbf{per candidate flow} & \cellcolor{lime} able to detect &\cellcolor{lime} robust \\
        \hline 
    \end{tabular}
    \label{tab:comparison}
\end{table}

Second, for a flow classifier trained on a labeled dataset of known flow types such as applications (\eg video streaming or conferencing) and content providers (\eg YouTube or Zoom), processing flows that do not belong to any known types is unavoidable in practical deployments. Although classifying flows into a finite coarse-grained scope can bypass unknown flow types, such as annotating flows by their network protocols and port numbers \cite{minzhao_cloud_gaming}, the value of classification results diminishes for large network operators that require visibility into trending applications and popular providers for network optimization and business strategies. Therefore, flow classifiers are expected to produce fine-grained application or provider-level classification while being capable of \textbf{detecting unknown flow types} rather than mislabeling them as one of the pre-defined known types. This is known as out-of-distribution classification in the machine learning community for scenarios in which it is unfeasible to enumerate all possible types. However, as listed in the third column of Table~\ref{tab:comparison}, such capability does not always exist due to their classifier architectures.

Third, packet sequence disorders in a flow caused by packet drops and retransmissions are commonly observed in large networks, where network conditions may not be ideal for every flow, such as the client devices served by lossy wireless connections, network congestion or bottlenecks on the routing path, or overwhelmed servers dropping
packets. Therefore, flow classifiers are expected to be reasonably \textbf{robust} to deviated time-series flow patterns caused by \textbf{packet sequence disorders}. As shown in the last column of Table~\ref{tab:comparison} and will be experimentally evaluated in \S\ref{sec:Evaluation}, prior works that precisely match a fixed number of packets for flow classification are not inherently robust to packet sequence disorders. Some prior works are arguably robust as they use aggregated statistics over a relatively lengthy interval (\eg per 40.96 seconds \cite{prototypical_networks_paper} or the entire duration of a flow \cite{unknown_flow_detection_clustering}), however, this inevitably sacrifices the speed of real-time flow classification.

\begin{figure}[t!]
    \centering
    \includegraphics[width= 0.99\linewidth]{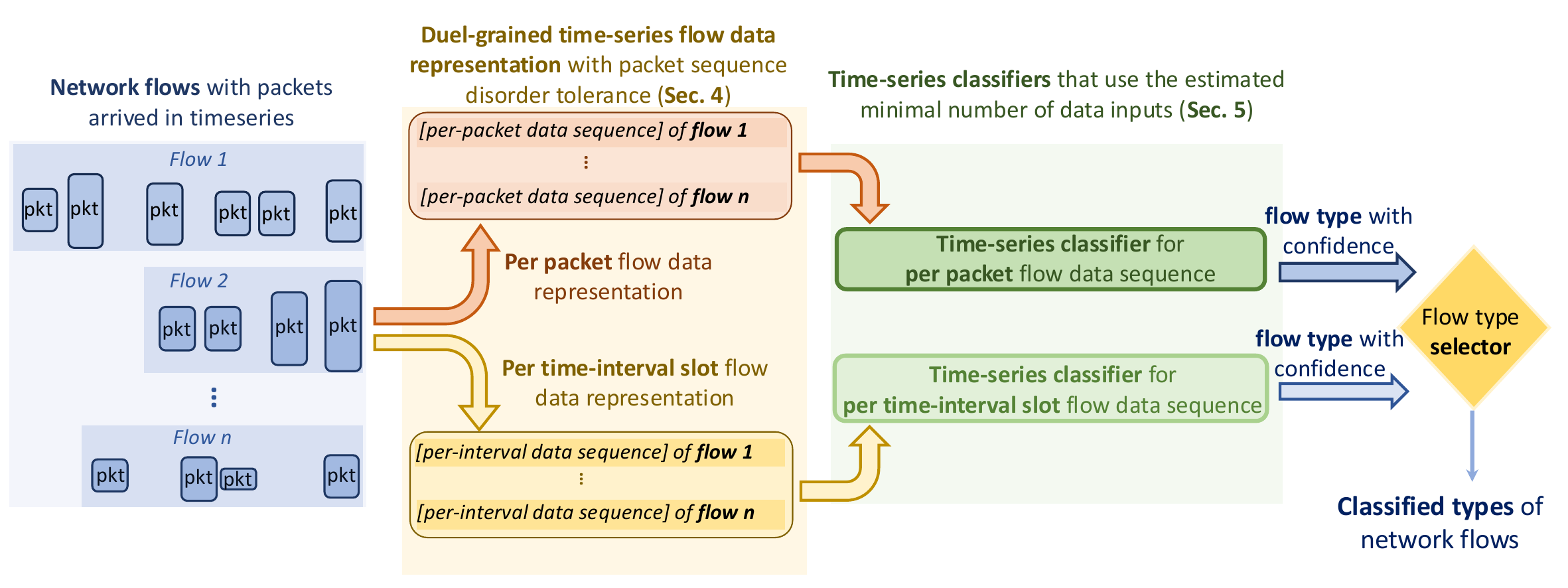}
    \vspace{-5mm}
    \caption{Overview of our \textit{FastFlow} flow classification process for large networks.}
    \label{fig:overview}
    \vspace{-5mm}
\end{figure}

\subsection{Overview of the \textit{FastFlow} Flow Classification Method}\label{sec:method}
Aimed to develop a flow classification method that addresses the three requirements for deployment in large networks, we leverage a combination of fine-grained packet sequence with coarse-grained slot sequence to represent time-series statistics per candidate flow. Our approach also avails of time-series classifiers trained with reinforcement learning for dynamic estimations on the minimal number of data points to take before making a confident classification per flow.
An overview of our method is provided in Fig.~\ref{fig:overview}, showing the steps starting with a flow represented as time-series data sequence until the final classification with confidence score. Our method aims to classify the real-time flows carried by a large network into known flow types (\ie types labeled during the training process) or the unknown flow type (\ie flow types not present in the training data) that are not of interest to a network operator or need further analysis.

Starting from the left side of Fig.~\ref{fig:overview}, all candidate flows that continuously have their packets flowing through the network will be represented as a time-series sequence containing statistics of the individual packets (\ie packet level) and aggregated per a given slot (\ie time interval). As will be detailed in \S\ref{sec:Representation}, this design takes advantage of both the precise flow statistics provided by packet sequence under ideal network conditions and the robustness to packet sequence disorders introduced by interval-based aggregation.
As shown in the green modules in Fig.~\ref{fig:overview} and discussed in \S\ref{sec:trainingProcess}, time-series flow data from each type of representation are fed into their respective time-series classifiers. In contrast to prior works in flow classification, each classifier in our design is trained with reinforcement learning techniques instead of supervised learning so that they are capable of making confident classification when sufficient (least number of) time-series inputs are provided for each candidate flow and identifying flows that do not belong to the known types.
Two different classifiers on packet and time interval real-time sequence data can produce unsynchronized results. For example, a flow with packet drops may not give a confident classification result until a large number of packets are received for distinguishable sequential patterns. In contrast, the data sequence aggregated by time interval can give fast prediction results with high confidence. To produce a fast and robust classification result per flow, a high-confident classification result from either one of the classifiers will first be selected.

\section{Precise yet Packet Sequence Disorder Robust Time-Series Flow Data Representation}\label{sec:Representation}
In this section, we discuss our dual-grained time-series data representation of a network flow that precisely preserve the fine-grained data sequence at (i) packet-level (\S\ref{sec:packetRepresentation}) for accurate classification without the occurrence of packet sequence disorders, assisted with (ii) coarse-grained data sequence aggregated per slot (\ie time interval) that is robust to the impact of packet sequence disorders on time-series classification performance (\S\ref{sec:timeIntervalRepresentation}). The classification result for a candidate flow at both granularities is selected by jointly considering their decision time and confidence \S\ref{sec:combination} for the earliest possible yet reliable decision.

\subsection{Time-Series Flow Data Sequence at Packet Granularity}\label{sec:packetRepresentation}
Under good conditions, network flows are expected to have lossless packet behaviors without disorder in their packet sequences caused by packet drops and retransmissions. As discussed by prior works \cite{ggfast,n_print,effective_packets,early_classification_application_negotiation}, network flows of a certain type, such as application types like video streaming or online gaming, application titles like different online games, or different user device operating systems (\eg iOS and Android), often exhibit unique sequence in their packet directions, sizes, and inter-arrival times. As revealed by prior works \cite{zoom_IMC_passive_2022,lyu2023metavradar,knowThyLag}, this is particularly true for their initial packets that carry static content (\eg requested services or application metadata) instead of those following ones depending on user's actions.

Therefore, to precisely preserve the packet-level statistics, we define our \textbf{fine-grained} time-series packet representation of a candidate flow as:

\vspace{-3mm}
\begin{equation}
\textit{\textbf{p}} = [\overrightarrow{p_1},\overrightarrow{p_2}, .., \overrightarrow{p_n}]
\label{formula:1}
\end{equation}

where as $\overrightarrow{p_n}$ describes statistics of the $n$th packet in this flow, including packet direction ($dir_n$), packet size ($s_n$), and inter-arrival time ($\Delta t_n$), which can be expressed as:

\vspace{-3mm}
\begin{equation}
    \overrightarrow{p_n} = \begin{pmatrix}
        dir_n & s_n & \Delta t_n       
    \end{pmatrix}^T
\end{equation}

\begin{wrapfigure}{l}{0.55\textwidth}
    \hspace{-4.5mm}
    \includegraphics[width= 1.1\linewidth]{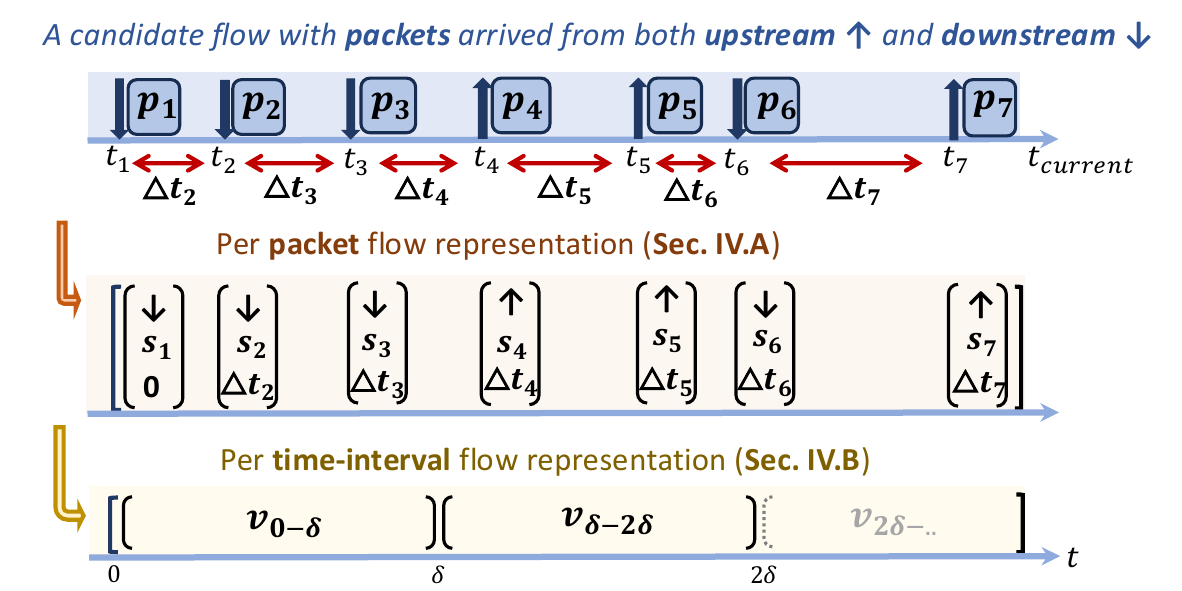}
    \vspace{-5mm}
    \caption{A candidate flow being represented as per packet and per slot time-series data sequence.}
    \label{fig:flowRepresentation}
\end{wrapfigure} 
A visual example of converting an ongoing flow into our per-packet time-series flow representation is provided in Fig.~\ref{fig:flowRepresentation}. From the middle layer of the figure, we can see that the statistics of every packet that has arrived until the current timestamp have been included in a runtime array for this candidate flow.
Our fine-grained time-series flow representation with packet attributes in matrix format is well compatible with popular time-series classification models such as Long-Short Term Memory Cell (LSTM) without extra overhead for pre-processing. Considering computational efficiency, in our later implementation, the packet direction $dir_n$ is set to 1 for upstream or 0 for downstream, the packet size $s_n$ is calculated as the packet payload size excluding the network and transport layer headers in the unit of MTU (\ie $s_n = payload_n / MTU$), and the packet inter-arrival time is taken as its log-scale conversion.  

\subsection{Time-Series Flow Data Sequence at Slot Granularity}\label{sec:timeIntervalRepresentation}
Our per packet sequence \textit{\textbf{p}} as defined in Equation~\ref{formula:1} works well for time-series flow classification when the packet-level profiles precisely match their expected patterns learned during the training process. However, this nearly ideal assumption might not always be realistic in deployment when the expected packet sequence of a candidate flow can be disordered due to packet drops and retransmissions caused by practical factors such as network, client, and server conditions. Therefore, we introduce our \textbf{coarse-grained} time-series representation for a candidate flow with each data point aggregated on slots, which can better handle packet sequence disorders caused by less ideal network conditions by trading off visibility into each packet.

We acknowledge that prior works \cite{Gaussian_model_for_fast_classification,sliding_window_N_packets} have also used aggregation per a certain number of packets instead of time intervals, as their classifications mainly use time-independent volumetric profiles such as average packet sizes. In early classification tasks, temporal information of a flow, such as the burstiness of the arrived packets, is equally important \cite{prototypical_networks_paper,lstm_10_20_30}. Therefore, we aggregate packets that fall into each time-interval slot by their arrival times to preserve both volumetric and temporal information.

Assuming that the first packet of a flow arrived at the timestamp $0$, and we have a time-interval slot of interest $\delta$ for aggregation, our slot time-series representation of a flow can be defined as:

\vspace{-3mm}
\begin{equation}
    \textit{\textbf{v}} = [\overrightarrow{v_{0-\delta}}, \overrightarrow{v_{\delta-2\delta}},\ldots, \overrightarrow{v_{(n-1)\delta-n\delta}}] 
\end{equation}

\noindent where $\overrightarrow{v_{(n-1)\delta-n\delta}}$ represents the statistics aggregated from the packets that arrived between the timestamps of $(n-1)\delta$ and $n\delta$.

\textbf{Aggregating packet statistics per time-interval slot:} We now describe our methodology for obtaining the aggregation data point $\overrightarrow{v_{(n-1)\delta-n\delta}}$. For simplicity, we use $\overrightarrow{v}$ for this data point, omitting its index in the rest of this section. We aim to explicitly preserve important contextual characteristics of a network flow that are inherently impacted by its functions and types \cite{Minzhao_IMC_User_platform,minzhao_cloud_gaming,lyu2023metavradar,zoom_IMC_passive_2022}, including packet directions, sizes of light and heavy packets, and dominant data transmission direction.
Also, the representation process should also be lightweight in real time, \ie can be achieved by online algorithms.
We therefore define the aggregated data point $\overrightarrow{v}$ in Equation~\ref{formula:4}:

\vspace{-3mm}
\begin{equation}
    \overrightarrow{v} = \begin{pmatrix}
        \overline{{\textbf{s}_{h\uparrow}}} &
        \overline{{\textbf{s}_{h\downarrow}}} &
        \overline{{\textbf{s}_{l\uparrow}}}&          \overline{{\textbf{s}_{l\downarrow}}} &
        \frac{\sum{\textbf{s}_{\uparrow}}}{\sum{\textbf{s}_{\downarrow}}} 
    \end{pmatrix}^T
\label{formula:4}
\end{equation}

\noindent where the first five items $\overline{{\textbf{s}_{h\uparrow}}}$, $\overline{{\textbf{s}_{h\downarrow}}}$, $\overline{{\textbf{s}_{l\uparrow}}}$, $\overline{{\textbf{s}_{l\downarrow}}}$ and $\frac{\sum{\textbf{s}_{\uparrow}}}{\sum{\textbf{s}_{\downarrow}}}$ denote the average packet payload size for upstream heavy packets, downstream heavy packets, upstream light packets, downstream light packets, and ratio between total upstream and downstream packet sizes per time-interval slot $\delta$, respectively. We use the average function for packet payload sizes as it is statistically important and can be computed using online algorithms in real-time. The light and heavy packet payload sizes are decided by a threshold value that can be set empirically for each deployment network. In our later implementation, we chose a threshold value of 1200 bytes to group packet payloads into light and heavy categories, which have been demonstrated as a reasonable separation.

\RestyleAlgo{ruled}
\SetNlSty{textbf}{(}{)}
\begin{algorithm}[t!]
    \fontsize{7}{9}\selectfont
  \textbf{\textit{Input}:} ($\textit{\textbf{class}}_p$, $\textit{\textbf{conf}}_p$) $\leftarrow$ classification results with confidence scores on packet flow data sequence, 
  ($\textit{\textbf{class}}_t$, $\textit{\textbf{conf}}_t$) $\leftarrow$ results on slot flow data sequence, 
  ($T_p$, $T_t$) $\leftarrow$ confidence thresholds for packet and slot flow data sequence, $\Delta select$ $\leftarrow$ user-defined selection time window;
  
  \textbf{\textit{Output}:} ($class_s$, $conf_s$) $\leftarrow$ the selected flow classification result and its confidence score;

  \textit{\textbf{Real-time flow classification result selection process:}}

   \tcp{\color{magenta}asynchronous event: a new result from the packet sequence classifier} 
   \nl \If{an arrived ($class_p$, $conf_p$) \textit{\textbf{and}} no ($class_t$, $conf_t$) arrive within $\Delta select$}{
    \eIf{$conf_p > T_p$}{
    \textbf{\textit{return}} ($class_s$, $conf_s$) $\leftarrow$ 
  ($class_p$, $conf_p$) \tcp{\color{magenta}confident flow type selected and exit}
    }{  \textit{\textbf{wait} for another data point to arrive}}
      } 
 
  \tcp{\color{brown}asynchronous event: a new result from the slot sequence classifier}
   \nl  \If{an arrived ($class_t$, $conf_t$) \textit{\textbf{and}} no ($class_p$,$conf_p$) arrive within $\Delta select$}{
     \eIf{$conf_t > T_t$}{
         \textbf{\textit{return}} ($class_s$, $conf_s$) $\leftarrow$ ($class_t$, $conf_t$) \tcp{\color{brown}confident flow type selected and exit}
     }{\textit{\textbf{wait} for another result to arrive}}
     }

  \tcp{\color{cyan}synchronous event: new results from both classifiers}
    \nl   \If{a ($class_p$, $conf_p$) \textit{\textbf{and}} a ($class_t$, $conf_t$) arrive within $\Delta select$}{
       \eIf{$conf_p > conf_t$ \textit{\textbf{and}} $conf_p > T_p$}{
       \textbf{\textit{return}} ($class_s$, $conf_s$) $\leftarrow$ ($class_p$, $conf_p$) \tcp{\color{cyan}flow type by packet sequence selected and exit}
       }{\textit{\textbf{wait} for another result to arrive}}
       \eIf{$conf_t > conf_p$ \textit{\textbf{and}} $conf_t > T_t$}{
       \textbf{\textit{return}} ($class_s$, $conf_s$) $\leftarrow$ ($class_t$, $conf_t$) \tcp{\color{cyan}flow type by slot sequence selected and exit}
       }{\textit{\textbf{wait} for another result to arrive}}
       }
  \caption{Selecting flow classification results produced by packet and slot classifiers nearly synchronously or asynchronously.}
  \label{algo:1}
\end{algorithm}

\subsection{Selecting Flow Classification Results at Real-Time}\label{sec:combination}

In our flow classification process provided in Fig.~\ref{fig:overview}, time-series data sequence of a candidate flow at the granularities of both packet (\S\ref{sec:packetRepresentation}) and slot (\S\ref{sec:timeIntervalRepresentation}) are fed into their respective classifiers, each generates a classification result (\ie flow type) with classification confidence. Our classifiers (to be discussed in \S\ref{sec:LSTMModel}) that work on packet or slot time-series data sequences can process their confident results for one candidate flow \textbf{asynchronously}. For example, a flow without packet sequence disorders can be confidently classified by the packet sequence classifier when the fifth packet arrives at a 0.5-second timestamp. In contrast, the slot classifier may produce its results at a 3-second timestamp. Alternatively, a flow with packet sequence disorders may never be confidently classified by its packet sequence classifier, while the slot classifier may generate a confident result within several seconds. In addition, both classifiers can produce their confident results without noticeable time differences, \ie \textbf{synchronously}. 
Therefore, to select the most accurate classification label for a candidate flow at the earliest possible timestamp, we design a real-time algorithm (algorithm~\ref{algo:1}) to select the confident classification result for a candidate flow from its labels produced by packet and slot classifiers asynchronously or synchronously.

The algorithm takes flow classification results and their confidence scores that are generated in real-time from both packet and slot sequence classifiers. As will be discussed soon in \S\ref{sec:LSTMModel}, the packet sequence classifier generates one result when a new packet arrives, and the slot classifier generates one prediction per time interval. Those results generated at runtime are continuously checked against preset confidence thresholds $T_p$ and $T_t$ to select a highly confident flow type. In our implementation, we use 90th-percentile of all confidence values obtained during the training process. We also provide a practical selection time window $\Delta select$ to recognize synchronously and asynchronously generated results by both classifiers.

For the asynchronous event, \ie a flow classification result is generated by one classifier, and no result is generated by the other one within $\Delta select$, as specified by blocks \textbf{(1)} and \textbf{(2)} in Algorithm~\ref{algo:1}, the classification confidence is checked with the respective threshold for the decision to accept the predicted flow type or wait for more results to come.
For the synchronous event processed by block \textbf{(3)}, the results of both classifiers are first compared with each other before checking with their respective confidence thresholds for a decision. In our engineering implementation, a bonus confidence score will be added to the classification result if both classifiers predict the same flow type.

\section{Time-Series Flow Classifier with the Estimated Minimal Length of Data Sequence}\label{sec:TimeSeriesModel}

After discussing how each network flow is represented as a time-series data sequence at both packet level (for precision) and slot level (for robustness to packet disorder), we now present our time-series (LSTM-based) classifiers (\S\ref{sec:LSTMModel}) that dynamically estimate the minimal length of data sequence required for accurate classification. Our LSTM classifiers are fitted with a reinforcement learning technique (\S\ref{sec:reinforcement}) on augmented labeled training datasets (\S\ref{sec:trainingProcess}).

\subsection{Time-Series Classifier Architecture for Early Flow Classification}\label{sec:LSTMModel}

Our approach uses time-series classifiers (green modules in Fig.~\ref{fig:overview}) to predict each flow type by its per packet or slot data sequences. After deployment, these classifiers are expected to process a large number of concurrent flows timely and accurately. An approach to increase the efficiency of the classifiers is to provide early predictions using a small number of time-series inputs (\ie packets or slots) for each flow. There is no single optimal number of time-series inputs for all flows, and a one-fits-all approach will not provide optimal results in practice.

\begin{figure}[t!]
    \centering
    \includegraphics[width= \linewidth]{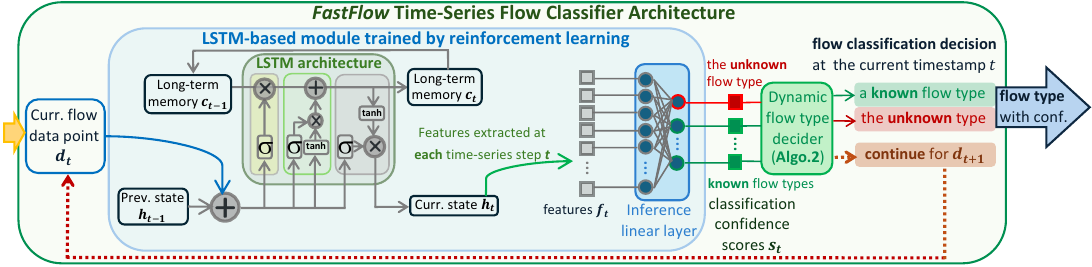}
    \vspace{-5mm}
    \caption{The architecture of \textit{FastFlow} time-series flow classifier realizing a sequential decision-making process.}
    \label{fig:classifier_plus_inference}
\end{figure}

However, existing works in flow classification typically train their models with a standard supervised approach that requires a fixed number of timesteps~\cite{lstm_10_20_30}, \cite{flow_pic}, \cite{CNN_LSTM_2}. The number of timesteps is considered a model hyperparameter optimized during training. Therefore, these methods will process the same amount of input for every flow and cannot adapt to the requirements of each candidate flow. Since the optimal time-series input length required for each flow may vary according to the classification objective and minor variations in the data sequence caused by network conditions, these approaches leave room for optimization. 
Moreover, traditional classifiers make a ``closed-world'' assumption that, after deployment, they will only observe the class labels present in the training set. However, this is an unrealistic assumption given the number and diversity of flows in operational networks. Consequently, these classifiers are doomed to misclassify unknown flows, unseen in the training data, as one of the known flow types~\cite{lessons_learned_from_commercial}. 

Existing works on flow classification using time-series models (especially deep-learning based) are prone to misclassify unknown flow types into known types \cite{cats_are_not_fish}. Even existing approaches that can address the open-world assumption often come with additional requirements such as additional data \cite{robust_network_traffic_classification}, extra computing power \cite{lessons_learned_from_commercial} or constraints on data distribution \cite{prototypical_networks_paper}.
Our approach addresses both requirements with a uniquely elegant solution: a \textit{sequential decision maker}. Under this paradigm, a classifier can take one of two decisions after receiving a new time-series data point: output a prediction or wait for the next data point. This decision is based on the classifier's confidence until the current data point. This approach has two major benefits: (i) the classifier issues a prediction as soon as it has gained enough confidence, and (ii) if no prediction is issued after enough data points, the classifier can issue an \textit{unknown} class label output.

Specifically, Fig.~\ref{fig:classifier_plus_inference} illustrates the architecture of our classifier. The blue rectangle indicates the section of the model trained with reinforcement learning. An LSTM cell provides a feature set at each time step. These features are fed into a fully connected network that outputs a class or an unknown label. The next section provides details on the training process for our trainable classifier module, which uses reinforcement learning instead of supervised learning approaches.

Our classifier leverages an LSTM \cite{LSTM_Paper} architecture due to its ability to process sequential data. Traditional LSTM-based classifiers extract features from a fixed number of timesteps and feed these features to a softmax layer to obtain scores for each known class label. As shown in the blue region of Fig.~\ref{fig:classifier_plus_inference}, our classifier module has two key design choices that differ from prior works. First, instead of waiting for a fixed number of timesteps, our model extracts features from the LSTM architecture every time a new data point arrives, which are then used by an inference linear layer for the classification confidence scores of each candidate class. Second, we introduce the `\textit{unknown}' flow type in addition to the existing types. Therefore, given a new flow, our model can output a temporary \textit{unknown} label for the initial timesteps and wait for more data points of the same flow until a confidence classification can be concluded or the maximum number of timesteps has been reached. The run-time decision on each time step is made by the dynamic flow type decider (green box in Fig.~\ref{fig:classifier_plus_inference}) as detailed in Algo.~\ref{algo:dynamicInference}.

\RestyleAlgo{ruled}
\begin{algorithm}[t!]
    \fontsize{7}{9}\selectfont
\SetKwInOut{Input}{input}  
\textbf{\textit{Input}:} Confidence scores of flow types $\vec{s_t}$ from the classifier, after processing the $t^{th}$ data point in the flow data sequence; confidence threshold $T_{unk}$; and the maximum time steps to make prediction $C_{unk}$.

\textbf{\textit{Output}:} Predicted flow type $l$ and the prediction confidence $p$.
    $\vec{c_t} \gets softmax(\vec{s}_t$)\\
    \If{$t = C_{unk}$}{
    \tcp{\color{red} End flow classification as the maximum number of time step $C_{unk}$ has been reached.}
        $l \gets$ `unknown'; \\
        $p \gets \vec{c}_t$[`unknown']    \\
        \Return{$l$, $p$}\\
    }

    \eIf{$\vec{c}_t[$`unknown'$] \geq T_{unk}$ or $argmax(\vec{c_t}) =~$`unknown'}{

        \tcp{\color{orange}Not yet confident enough to make a `not unknown' prediction, wait for $\vec{s}_{t+1}$}
        $l \gets$ `unknown' \\
        $p \gets \vec{c}_t$[`unknown'] \\
        \textbf{continue} for the next time step $t+1$.

    }{  
        \tcp{\color{blue}Confident enough to make a `not unknown' prediction}
        $l \gets argmax(\vec{c}_t)$ \\
        $p \gets \vec{c}_t[l]$ \\
        \Return{$l$, $p$}\\
    }
  \caption{Dynamic inference on time-series flow data sequence.}
  \label{algo:dynamicInference}
\end{algorithm}

The algorithm processes the confidence of the $k$ known classes plus the extra unknown class for each incoming time-series data point. If the confidence of the unknown class is greater than the threshold $T_{unk}$ or if it is the most confident class, a temporary unknown classification is decided, making the classifier continue for the next time series data point. Otherwise, the algorithm produces the most confident class among the $k$ known classes. This process continues until an output is issued or the classifier processes a maximum of $C_{unk}$ timesteps. We note that both thresholds, $T_{unk}$ and $C_{unk}$, are hyperparameters. In our training process, we tune $C_{unk}$ to provide a balance between prediction accuracy and speed.

\subsection{Training \textit{FastFlow} Classifiers with Reinforcement Learning} 
\label{sec:reinforcement}

We have conceived the network flow classification task as a sequential decision problem, making traditional supervised learning methods unsuitable for training our models. Traditional supervised learning defines a loss function computed for every training case. Therefore, these methods assume that every case should immediately lead to some loss after classification. However, our design introduces a postponing decision that does not immediately lead to a loss of value.

Reinforcement learning (RL)~\cite{reinforcement_learning} is a better-suited approach to train our models. RL involves an agent (\ie sequential decision-making classifier) that observes the current environment state and chooses to take an action from a predefined action space. Based on this action, the environment will update its current state. Each time the agent/classifier selects an action, it gets a reward based on its fit to the current state. This process repeats until the environment arrives at a terminal state, where the agent cannot change the environment's state further. RL learning aims to maximize the total reward granted to the agent from the start state to the terminal state. 

There are many popular RL algorithms, such as Q-learning \cite{Q_learning}, PPO \cite{PPO}, and A3C \cite{A3C}. We employ Q-learning since our flow classification leads to a discreet action space. Q-learning is also known to be sample efficient~\cite{q_learning_sample_effecient}, which is crucial when collecting labeled data is expensive. Finally, Q-learning had several enhancements over the initial algorithm, such as model architecture structure~\cite{dueling}, loss function~\cite{double_q_learning}, and training process~\cite{pQ_learning}, which have improved its performance, as our experiments confirm.
Q-learning trains the agent to choose the best action from the action space given a state. Given a current state $s$ as input, the agent outputs a number (Q value) for each action potential $a \in A$ in the action space $A$. The Q-value $Q(s, a)$ represents the expected reward for taking action $a$ in state $s$. Therefore, after training, the agent always picks the action with the maximum Q-value given an input state.

In our flow classification approach, the agent is our classifier $\mathcal{C}$. The state is the first $p$ timesteps of the flow representation (in either packet or slot). The action space for a $k$-way classification problem has $k+1$ actions, one for each flow type, and the $k+1th$ action for the unknown class. The terminal state occurs when the agent chooses to make a prediction or when it has seen a maximum number of timesteps, $C_{unk}$. The reward given to the agent is decided by the reward function that takes the current state $s$, the action $a$, and the actual class label $l$ of the current flow. 

\vspace{-3.5mm}
\begin{equation}
    \mathbf{reward\_func}(s,a,l) = 
\begin{cases}
     wait\_penalty &  \text{if    }      a == k+1 \\
    positive\_reward   &   \text{else if  } a == l\\
    negative\_reward   & \text{otherwise}
\end{cases}
\label{func:reward_function}
\end{equation}

The reward function outputs a positive reward (`$+1$' in our training process) if the classifier outputs a correct prediction; or a negative reward (`$-1$' in our implementation) for an incorrect prediction. However, if the classifier decides to output `unknown' and continues for the next timestep, a negative reward `$wait\_penalty$', a hyperparameter that determines the speed of classification, will be awarded. We set the value of `$wait\_penalty$' as `$-.03$' in our experiments after iterative selections. Lowering its value will place more importance on classification speed than accuracy and vice versa. 
The model is then trained using double Q learning loss with sampling priority \cite{double_q_learning,pQ_learning} that inputs tuples of $(state,action, next\_state,reward)$.

\subsection{Training Flow Data Augmentation}\label{sec:trainingProcess}
A final component of our solution involves the generation of flows with unknown labels. In many real scenarios, these unknown flows are readily available, consisting of all flows that do not belong to the interest classes. However, we may find some challenges in collecting these flows in practice. One situation occurs when these flows need to be manually labeled to guarantee they do not fall into one of the existing classes. Another practical situation involves training models using benchmark datasets, as most of these datasets do not bring a general unknown class.

A solution to this problem is to generate synthetic unknown flows using augmentation techniques. Augmentation is a widely used approach in deep learning to solve data deficiency~\cite{data_augmentation}, as several deep models have a large number of parameters that require extensive training sets. We use a strong augmentation approach that generates synthetic unknown flows away from the high-density areas that characterize each known flow type. 

\begin{wrapfigure}{l}{0.6\textwidth}
    \centering
        \begin{subfigure}[]{0.323\linewidth}
        \centering
        \includegraphics[width=\linewidth]{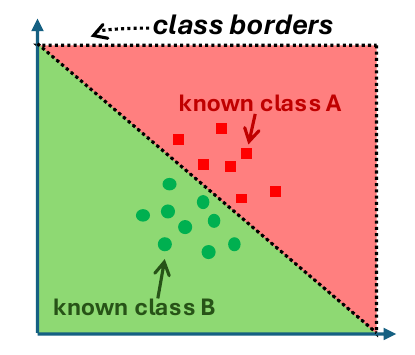}
        \caption{Classifier trained on known classes.}
        \label{fig:closed_training}
    \end{subfigure}
    \begin{subfigure}[]{0.323\linewidth}
        \centering
        \includegraphics[width=\linewidth]{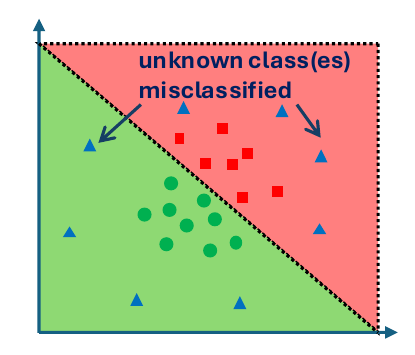}
        \caption{Misclassifying unknown class(es).}
        \label{fig:closed_testing}
    \end{subfigure}
    \begin{subfigure}[]{0.323\linewidth}
        \centering
        \includegraphics[width=\linewidth]{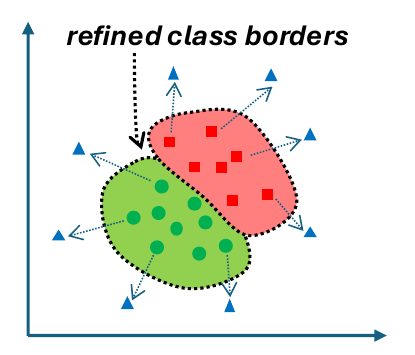}
        \caption{Classifier with refined class borders}
        \label{fig:closed_augmented}
    \end{subfigure}
   \vspace{-4mm}
    \caption{(\textit{a}) Under the closed-world assumption, a classifier can naively split the entire feature space according to the existing classes. (\textit{b}) The resulting classifier misclassifies unknown instances as belonging to one of the existing classes. (\textit{c}) We use augmentation in classifier training to generate closely outlying synthetic unknown flows that help bind the classes' decision borders.}
    \label{fig:closed-open-world}
\end{wrapfigure} 
Fig.~\ref{fig:closed-open-world} is a simplified example that illustrates using strongly augmented flows to restrain the decision border around each class, so that the trained classifier is capable of detecting flow instances belonging to the unknown types (\ie not statistically belonging to any known class) with the refined borders of each flow type.

Specifically, this synthetic unknown flow augmentation is achieved by distorting the attributes of packets before converting them into packet or slot flow data representation. 
We start by sampling $\alpha_{attr} \in [0,1]$ fraction of packets in each flow. The payload size, $ps_i$, direction, $dir_i$, and timestamp, $t_i$ for each sampled $i$th packet are then distorted as follows.

\vspace{-3mm}
\begin{align}
    &ps_i^{aug} = ps_i*\alpha_{ps} + \text{U}(0, MTU)*(1 - \alpha_{ps}) \\
    &t_i^{aug} = t_i + ((t_{i+1} - t_{i})*\text{U}(0,1) - (t_i - t_{i-1})*\text{U}(0,1)) *\alpha_{ts}
\end{align}

\noindent where $\alpha_{ps} \in [0,1]$ and $\alpha_{ts} \in [0,1]$ are the strength of payload and timestamp distortion, respectively. $U(a,b)$ denotes the uniform distribution over the range $[a,b]$. To apply augmentation for the attribute $dir_i$, the direction of a flow is randomly reversed with probability $\alpha_{dir}$. In practice, we set $\alpha_{ps}, \alpha_{ts}$ and $\alpha_{dir}$ to `$0.2$' after experimental selections. $\alpha_{attr}$ is sampled from $U(0.6,0.9)$ to generate highly augmented pseudo unknown flows. We represent this augmentation over attributes using $\mathcal{F}_{attr}$, as formally defined in formula~\ref{formula:augmentation}.

\vspace{-3mm}
\begin{equation}
\mathbf{augmented\_training\_flow} = \mathcal{F}_{\text{attr}}(\textit{flow\_{known}}, \alpha_{\text{attr}}, \alpha_{ps}, \alpha_{ts}, \alpha_{\text{dir}})
\label{formula:augmentation}
\end{equation}

Notably, the pseudo unknown flows used in the classifier training process will be labeled `\textit{unknown}'. Therefore, during training, the classifier will always get a negative reward when it misclassifies these flows into one of the known types. This enhances the unknown flow detection capability of our trained classifier. We also create a weaker form of augmentation by setting $\alpha_{attr}$ from $U(0,0.2)$ to increase the size of the training data and mitigate the class imbalance \cite{augementation_for_class_imbalance}, which are also tested as effective in our evaluation using public datasets and campus network deployment.

\section{Evaluation, Benchmarking and Deployment Insights}\label{sec:Evaluation}

This section thoroughly evaluates \textit{FastFlow} for flow classification, focusing on classification performance (accuracy and execution time) and recognition of unknown flows in line with our design objectives. This section starts by describing the experimental setup used in our experiments (\S\ref{sec:experimental_setup}), followed by a comparison with the state-of-the-art (\S\ref{sec:evalAndComparison}). We also provide an ablation study evaluating the necessity of our use of packet and slot flow data sequence (\S\ref{sec:ablationStudy}), and conclude this section with deployment insights in our university campus network that mimics a residential ISP (\S\ref{sec:campusDeployment}).

\subsection{Lab Evaluation Setup}\label{sec:experimental_setup}

Our lab experiments evaluate the performance of \textit{FastFlow} in achieving two primary objectives: \textit{i}) classifying network flows accurately with fast speed through the use of the estimated minimal number of packets per flow and \textit{ii}) maintaining classification robustness to packet sequence disorders and unknown flow types commonly found in deployments. To achieve this objective we use three popular public datasets containing labeled packet captures (PCAPs) of network flows for various network types, namely VNAT \cite{prototypical_networks_paper} by MIT Lincoln lab for VPN networks, UTMobileNet \cite{UTMobileNet2021} for mobile networks, and UNIBS \cite{UNIBS_DATASET} collected from the campus network edge of the University of Brescia. Table~\ref{tab:combined_dataset_samples} lists the flow types in each dataset.

\begin{wraptable}{l}{5.6cm}
\vspace{-4mm}
\caption{Specification of the three public network flow datasets used in our evaluation.}\label{tab:combined_dataset_samples}
\vspace{-3mm}
\fontsize{7}{9}\selectfont
\begin{tabular}{|l l|l l|}
\hline
\multicolumn{2}{|c|}{\cellcolor[rgb]{ .906,  .902,  .902}\textbf{UTMobileNet}} & \multicolumn{2}{c|}{\cellcolor[rgb]{ .906,  .902,  .902}\textbf{UNIBS}} \\
\hline
\textbf{Flow type} & \textbf{\#flows} & \textbf{Flow type} & \textbf{\#flows} \\ \hline
Google-Maps & 2584 & Browsers & 18820 \\
Netflix & 1680 & P2P & 14175 \\
Reddit & 1295 & Mail & 4432 \\
Youtube & 1031 & Other & 2368 \\
Facebook & 1028 & Skype & 499 \\
Instagram & 994 & & \\
 \cline{3-4}
Pinterest & 994 &   \multicolumn{2}{|c|}{\cellcolor[rgb]{ .906,  .902,  .902}\textbf{VNAT}}  \\
\cline{3-4} 
Google-Drive & 709 &Streaming  &  1628\\
Spotify & 699 & Chat & 811 \\
Twitter & 682 & Control   & 611 \\
Gmail & 400 & File Transfer  & 456 \\
Hangout & 351 &   &  \\
Messenger & 329 &  &  \\ \hline
\end{tabular}
\vspace{-3mm}
\end{wraptable}

A common limitation of these public datasets is that they are collected under nearly ideal network conditions, with no packet drop or retransmission in each labeled flow. Also, they do not contain unknown (unlabelled) flows. Therefore, we augment these datasets to introduce packet sequence disorders and randomly exclude certain flow types to mimic unknown flows in our evaluation.

To simulate packet sequence disorder, we randomly drop $\alpha_{drop} \in [0,100)$ percentage of packets in the flow. For TCP flows, these packets are reinserted if the flow uses TCP after a variable retransmission delay. In practice, we set the retransmission delay as the RTT calculated by a three-way handshake. We chose the the drop probability $\alpha_{drop}$ of each packet randomly following a normal distribution with a mean value of 5\% and a standard deviation of 3.5\%, as suggested by the network operation community \cite{obkio_packet_drop,websentra_packet_drop}. This models a scenario with more realistic network conditions. 



To test the capability of detecting unknown flow types, we randomly exclude around 20-25\% of the flows from the training set for each training and evaluation iteration. More precisely, we remove three flow types from UTMobileNet, and one flow type from UNIBS and VNAT. We perform ten evaluation iterations for each dataset. We guarantee that every flow type is excluded from the training set at least once. During each evaluation iteration, we randomly use 70\% of the data for model training and the remaining 30\% data for testing.

We must clarify that these augmentations were chosen to give existing datasets more realistic data characteristics. These augmentations were \textit{not} introduced into the data to benefit \textit{FastFlow}, besides the fact that our method was designed to deal with packet drops and unknown flows. To certify that \textit{FastFlow} is not favored, the packet drops are \textit{only} introduced in the test data. Thus, neither \textit{FastFlow} nor its competitors have information about the packet drop rate during training. We also present results of the non-augmented dataset in Appendix \ref{sec:Appendix_Eval}, so the reader can have a complete picture of \textit{FastFlow} performance with all combinations of presence and absence of packet drops and unknown flows.

Our experiments use different performance measures. We report accuracy and macro F1 for identifying known flow types. Accuracy is a prevalent performance measure, but it is difficult to interpret in the presence of several flow types and when some are uncommon. Macro F1 is the unweighted average of the F1 measure for each flow. This measure favors classifiers that perform well for all flow types independently of the number of flows. Both measures are shown as percentages, with higher values indicating better classifiers.

We report the false positive rate (FPR) and the true positive rate (TPR) to assess the recognition of unknown flows. The FPR measures the percentage of unknown flows incorrectly assigned to one of the existing flow types. The TPR measures the percentage of unknown flows correctly recognized as such. Finally,  we report the number of packets and inference time in seconds as measures of inference efficiency. 

We conclude our experiment setup with some remarks about \textit{FastFlow} hyperparameters to enable the reproducibility of our results. The hyper-parameters for the \textit{FastFlow}' LSTM cells are set through standard tuning processes, including a hidden size of 128, Adam optimizer with a learning rate of $3e-4$, and a capped training epoch of 200. The static parameters in \textit{FastFlow} classifier architecture are empirically tuned for a balanced classification and speed per deployment network (or dataset). For example, the time-interval slot is configured as 50ms, $C_{unk}$ is set to 20, 30, and 15 for UNIBS, UTMobilenet and VNAT dataset, respectively. In our campus deployment, this value is set to 25 during the training process. $T_{unk}$ and $wait\_penalty$ are consistently set to 0.8 and -0.03, respectively.

\begin{table}[t!]
\centering
\caption{Classification performance comparison between \textit{FastFlow}, the state-of-the-art and baselines under packet sequence disorder and unknown flow types.}
\vspace{-3mm}
\label{tab:results_table_ood_packet_drop}
\fontsize{7}{9}\selectfont
\begin{tabular}{|l|l|c|c|c|c|c|c|}
\hline
\multirow{2}{*}{\textbf{Dataset}} & \multirow{2}{*}{\textbf{Method}}  & \multicolumn{4}{c|}{\cellcolor[rgb]{ .906,  .902,  .902}\textbf{Classification performance on known flow types}} & \multicolumn{2}{c|}{\cellcolor[rgb]{ .906,  .902,  .902}\textbf{Unknown types}} \\ \cline{3-8}
                                  &                                   & \textbf{Macro F1 (\%)} & \textbf{Accuracy (\%)} & \textbf{Packets (\#)} & \textbf{Time (s)} & \textbf{FPR (\%)} & \textbf{TPR (\%)} \\ \hline
\multirow{3}{*}{UTMobileNet} 
                             & {\color{blue}\textbf{FastFlow}}     & {\color{blue}\textbf{85.42}}   & {\color{blue}86.32} & {\color{blue}12.92 $\pm$ 9.48} & {\color{blue}0.57 $\pm$ 1.61} & {\color{blue}4.56} & {\color{blue}86.94}                \\ 
                             & {\color{purple}GGFast \cite{ggfast}} & {\color{purple}77.82} & {\color{purple}83.90} & {\color{purple}50} & {\color{purple}2.56 $\pm$ 0.49} & {\color{purple}\textbf{1.04}} & {\color{purple}\textbf{90.42}}  \\
                             & {\color{purple}Grad-BP \cite{lessons_learned_from_commercial}} & {\color{purple}80.33} & {\color{purple}83.85} & {\color{purple}100(TCP) 10(UDP)} & {\color{purple}4.86 $\pm$ 1.98} & {\color{purple}4.92} & {\color{purple}61.35} \\
                             &  Pkt.-5 & 65.28 & 74.20  & 5  & 0.28 $\pm$ 1.07 & -- & --  \\
                             & Pkt.-45 &  80.02  & 81.17 & 45 & 2.40 $\pm$ 1.24 &-- & -- \\
                             & Time-int.-5 & 73.39 & 75.96 & \textbf{4.39 $\pm$ 1.96} & \textbf{0.25}& -- & -- \\
                             & Time-int.-45 & 85.29 & \textbf{87.05}  & 42.02 $\pm$ 9.23 & 2.25 & -- & -- \\ \hline
\multirow{2}{*}{VNAT} 
                            & \textbf{\color{blue}FastFlow}     & {\color{blue}96.24} & {\color{blue}\textbf{98.03}} & {\color{blue}\textbf{4.16 $\pm$ 1.34}} & {\color{blue}\textbf{0.033 $\pm$ 1.43}} & {\color{blue}\textbf{0}} & {\color{blue}97.32}                \\ 
                            & {\color{purple}GGFast \cite{ggfast}} & {\color{purple}70.45} & {\color{purple}86.68}  & {\color{purple}50} & {\color{purple}1.08 $\pm$ 0.42} & {\color{purple}\textbf{0}} & {\color{purple}\textbf{99.47}} \\
                            & {\color{purple}Grad-BP \cite{lessons_learned_from_commercial}} & {\color{purple}95.64} & {\color{purple}95.91} & {\color{purple}100(TCP) 10(UDP)} & {\color{purple}1.91 $\pm$ 0.56} & {\color{purple}4.98} & {\color{purple}93.07} \\
                            &  Pkt.-5 & 63.53 & 71.48  & 5  & .035 $\pm$ 0.81 & -- & -- \\
                            & Pkt.-45 & 80.41 & 85.66 & 45 & 0.96 $\pm$ 1.59 & -- & -- \\
                            & Time-int.-5 & \textbf{96.48} & 97.59 & 14.22 $\pm$ 5.42 & 0.25 & -- & -- \\
                            & Time-int.-45 & 95.20 & 97.45 & 100.47 $\pm$ 23.44 & 2.25 & -- & -- \\ \hline
\multirow{2}{*}{UNIBS} 
                            & \textbf{\color{blue}FastFlow}     & {\color{blue}92.30}  & {\color{blue}95.21} & {\color{blue}9.92 $\pm$ 3.92} & {\color{blue}0.36 $\pm$ 0.82} & {\color{blue}2.54} & {\color{blue}98.16}                \\ 
                            & {\color{purple}GGFast \cite{ggfast}} & {\color{purple}87.93} & {\color{purple}91.95} & {\color{purple}50} & {\color{purple}1.52 $\pm$ 1.87} & {\color{purple}\textbf{0.87}} & {\color{purple}\textbf{99.37}} \\
                            & {\color{purple}Grad-BP \cite{lessons_learned_from_commercial}} & {\color{purple}90.15} & {\color{purple}93.46} & {\color{purple}100(TCP) 10(UDP)} & {\color{purple}2.95 $\pm$ 1.97} & {\color{purple}4.93} & {\color{purple}68.45} \\
                            & Pkt.-5 & 81.49 & 87.02 & \textbf{5} & 0.19 $\pm$ 1.18 & --  & -- \\
                            & Pkt.-45 & 90.53 & 94.45 & 45 & 1.36 $\pm$ 1.46 & -- & -- \\
                            & Time-int.-5 &81.91  & 82.84 & 6.83 $\pm$ 2.80 & \textbf{0.25}  & -- & -- \\
                            & Time-int.-45 & \textbf{92.36} & \textbf{95.55} & 75 $\pm$ 17.15 & 2.25& -- & -- \\ \hline
\end{tabular}
\end{table}

\subsection{Evaluation and Comparison to the State-of-the-art}\label{sec:evalAndComparison}

We evaluate the performance of \textit{FastFlow} and compare it with two representative state-of-the-art network flow classification methods recently developed by the research community: \textit{GGFast} \cite{ggfast} and \textit{Grad-BP} \cite{lessons_learned_from_commercial} as well as some baselines. Both \textit{GGFast} and \textit{Grad-BP} have been practically deployed in large networks, as discussed in their respective papers. \textit{GGFast} classify flows by their packet length sequences and \textit{Grad-BP} uses deep learning models to classify flow types by its packet lengths and directions in time-series sequences. The two methods are trained with their original design without our augmentation technique for unknown flow detection as discussed in \S\ref{sec:trainingProcess}.
Other promising works such as CATO \cite{CATO} for flow classification in large networks leveraging different approaches like multi-objective optimization, are not all extensively evaluated.

Table~\ref{tab:results_table_ood_packet_drop} summarizes our evaluation results\footnote{Additional results of our extensive evaluation are available in Appendix~\ref{sec:Appendix_Eval}. They include \textit{i}) ideal network flow conditions on the original datasets, \textit{ii}) with only packet sequence disorders, and \textit{iii}) with only unknown flow types. \textit{FastFlow} classifiers perform satisfactorily in all three scenarios.}
\textit{FastFlow} outperforms \textit{GGFast} and \textit{Grad-BP} in terms of classification performance for known flow types and inference efficiency. \textit{GGFast} performs better than \textit{FastFlow} for detecting unknown flows. However, \textit{GGFast}'s small decrease in FPR and increase in TPR for unknown flows comes with a significant decrease in performance for the classification of known flow types. For example, for the VNAT dataset, \textit{GGFast} macro F1 is only 70.45\% where \textit{FastFlow} archives a respected 96.24\%. Similar results can be observed for the other datasets.
Compared with \textit{FastFlow}, the performance drop of \textit{GGFast} may be due to its dependence on packet sequence patterns that can be affected by packet drops and retransmissions. For \textit{GradBP}, apart from its choice of using a fixed number of packets for flow classification, its performance in detecting unknown flows may be affected by not having a dataset augmentation technique for training the unknown flow detection capability.

\begin{figure*}[t!]
    \centering
    \begin{subfigure}[b]{0.23\linewidth}
        \centering
        \includegraphics[width=\linewidth]{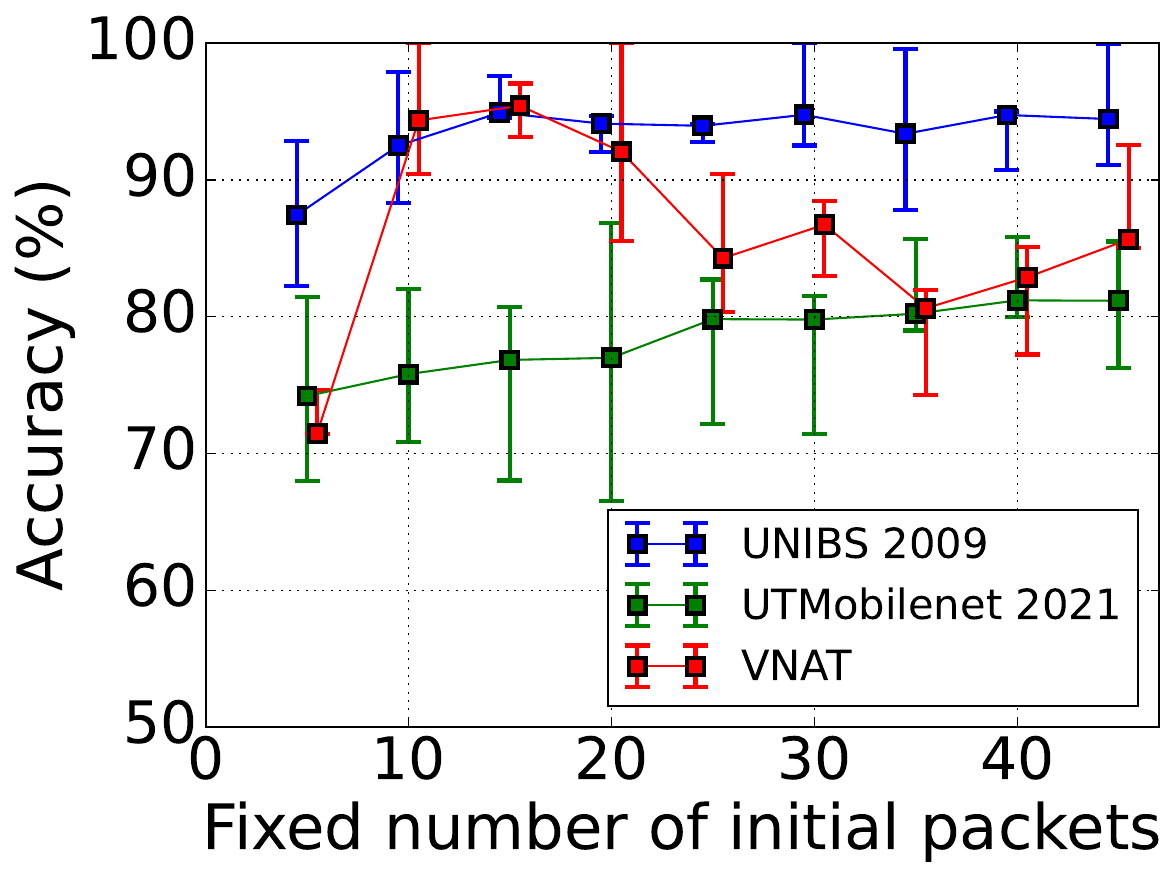}
            \caption{Accuracy for using a fixed number of packets.}
        \label{fig:FP_accuracy}
    \end{subfigure}
    \hfill
    \begin{subfigure}[b]{0.23\linewidth}
        \centering
        \includegraphics[width=\linewidth]{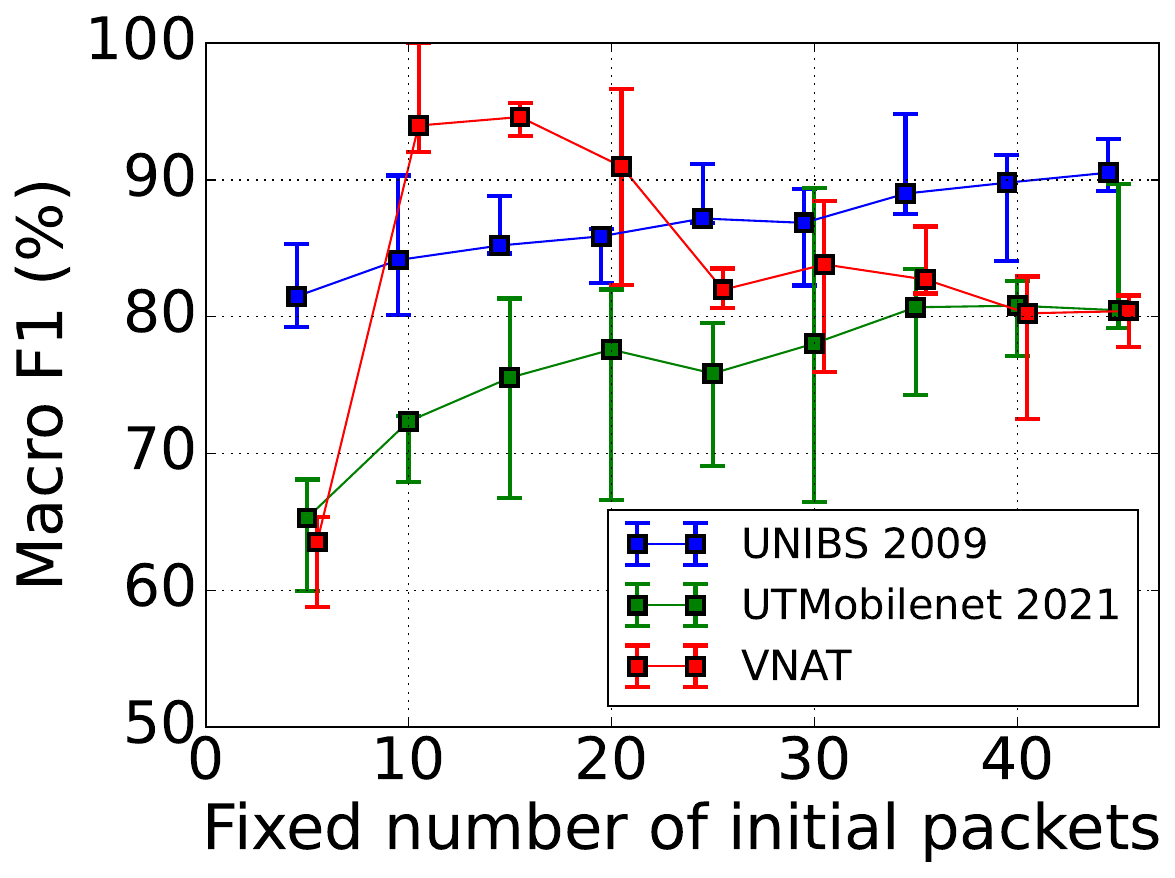}
        \caption{Marco F1 for using a fixed number of packets.}
        \label{fig:FP_F1}
    \end{subfigure}
    \hfill
    \begin{subfigure}[b]{0.23\linewidth}
        \centering
        \includegraphics[width=\linewidth]{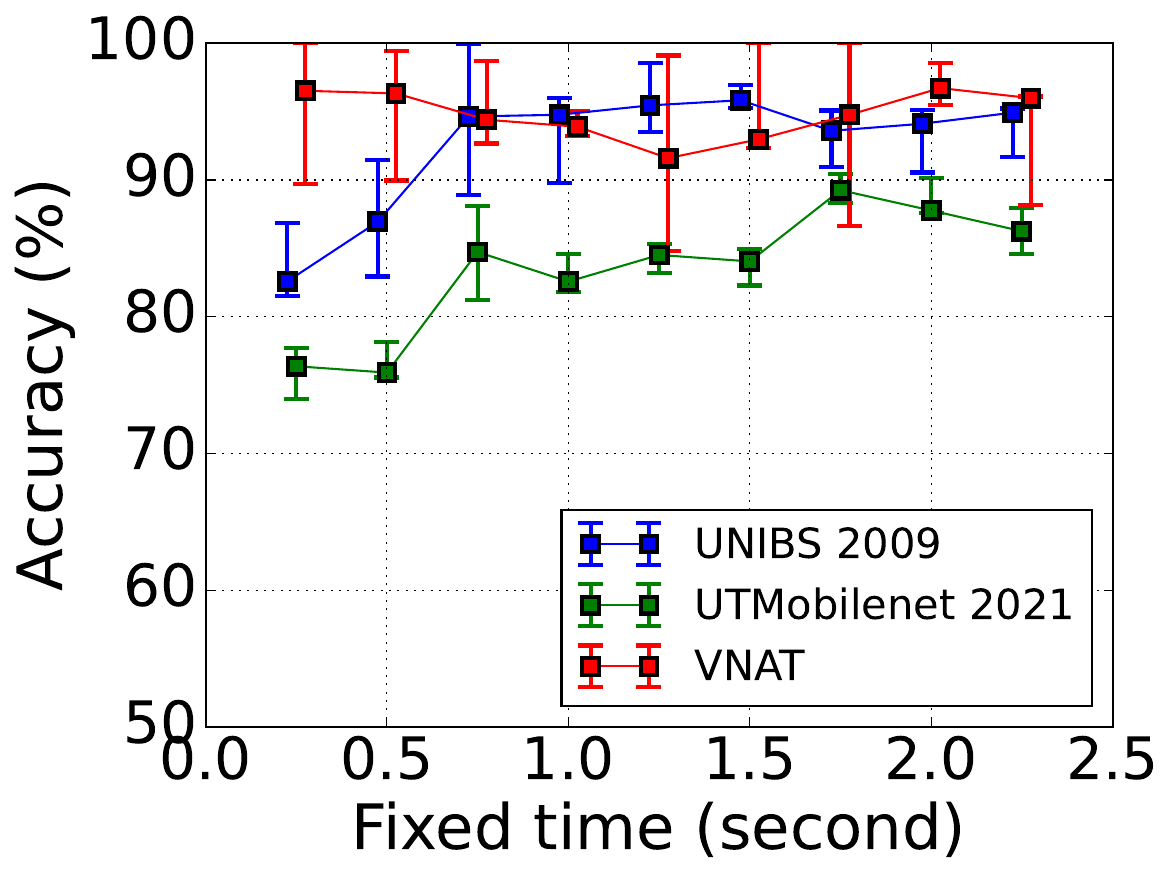}
        \caption{Accuracy for using fixed time-interval slots.}
        \label{fig:FT_accuracy}
    \end{subfigure}
    \hfill
    \begin{subfigure}[b]{0.23\linewidth}
        \centering
        \includegraphics[width=\linewidth]{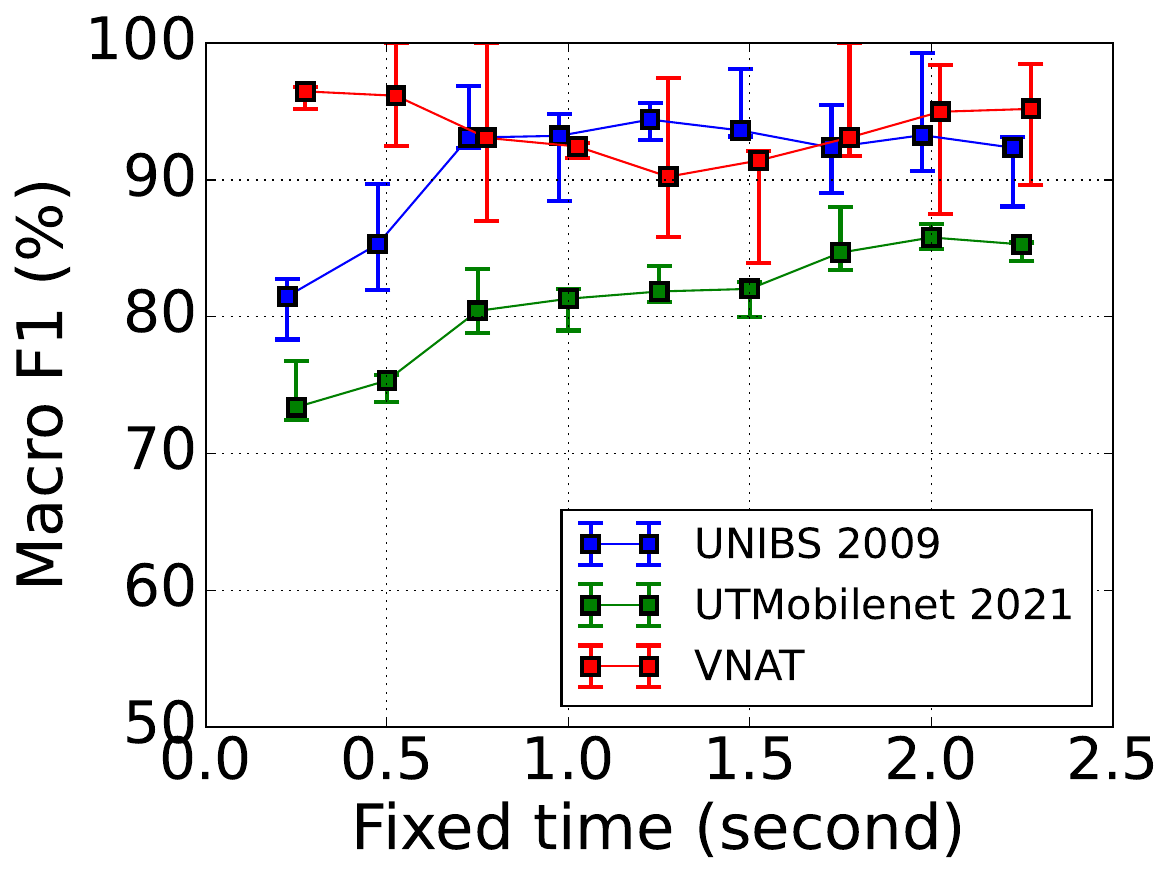}
        \caption{Marco F1 for using fixed time-interval slots.}
        \label{fig:FT_F1}
    \end{subfigure}
    \vspace{-3mm}
    \caption{Performance of classifiers when using time-series statistics from a fixed number of packets or time-interval slots on the three public datasets. The bounded range of each data point shows the variation of accuracy/marcoF1 across flow types in each dataset. }
    \label{fig:fp_ft_compare}
\end{figure*}

We also include some baselines that represent families of methods in the literature. For example, several prior works classify network flows using a fixed number of time-series data points of either packets or time-interval slots~\cite{ggfast, prototypical_networks_paper, lstm_10_20_30, Gaussian_model_for_fast_classification}. We include the performance of standard LSTM time-series classifiers that take a fixed number of initial flow data points without the ability to detect unknown flows, therefore, their respective values in Table~\ref{tab:results_table_ood_packet_drop} are marked as `-'. These classifiers use the first five packets (named `Pkt.-5' in Table ~\ref{tab:results_table_ood_packet_drop}), the first forty-five packets (named `Pkt.45'), the first five slots (named `Time-int.-5'), and the first forty-five slots (named `Time-int.-45'). 

\textit{FastFlow} outperforms most of the baselines. Pkt.-5 clearly does not have enough information to make precise classifications. Pkt.-45 shows that increasing the number of packets does improve classification performance but not enough to achieve \textit{FastFlow}'s performance. We did not investigate the performance of packet classifiers with more than 45 packets since the inference times for Pkt.-45 are already much larger than \textit{FastFlow}. Slot classifiers perform better than packet classifiers. Overall, five slots are not enough to guarantee good performance, but with 45 intervals, the classifiers can reach similar classification performance to \textit{FastFlow} and even slightly outperform our proposal. However, this comes with a longer inference time. We also notice that all baselines cannot handle unknown flows, so we cannot measure their performance on this task.

In addition to the two fixed numbers (\ie 5 and 45) of packets or slots for flow classification, we have also tested other fixed numbers from 5 to 45 with a step of 5. The results, including accuracy and marco F1 score, are averaged for each dataset and are provided in Fig.~\ref{fig:fp_ft_compare}. We can see that \textit{FastFlow} outperforms all settings with fixed number of inputs in both accuracy and marco F1 score. From Fig.~\ref{fig:fp_ft_compare}, it can also be observed that using more initial packets or time-interval slots to classify a flow does not necessarily produce better accuracy, validating the merit of \textit{FastFlow} method that can dynamically estimate the minimal number of packets or slots to classify each candidate flow.

\subsection{Ablation Study with Only Packet or Slot Flow Data Sequence}\label{sec:ablationStudy}

This section presents additional analyses that provide a better understanding of \textit{FastFlow} performance. We start with an ablation study that characterizes the proposal's performance with either packet or slot representation.

\textit{FastFlow} collectively uses two data representations: packet and slot data sequence. We analyze the importance of using both representations instead of one alone. This study excludes one time-series classifier (green boxes in Fig.~\ref{fig:overview}) trained over one flow data representation (red or yellow boxes in the same figure). Table~\ref{tab:ablation_results_table_ood_packet_drop} summarizes the results, where \textit{Packet seq.} stands for a \textit{FastFlow} version using only a packet representation and \textit{Time-int. seq.} using only a slot representation.

\begin{table}[t!]
\centering
\caption{Ablation study comparing \textit{FastFlow} with only one input flow data representation.}
\vspace{-3mm}
\label{tab:ablation_results_table_ood_packet_drop}
\fontsize{7}{9}\selectfont
\begin{tabular}{|l|l|c|c|c|c|c|c|}
\hline
\multirow{2}{*}{\textbf{Dataset}} & \multirow{2}{*}{\textbf{Method}}  & \multicolumn{4}{c|}{\cellcolor[rgb]{ .906,  .902,  .902}\textbf{Classification performance on known flow types}} & \multicolumn{2}{c|}{\cellcolor[rgb]{ .906,  .902,  .902}\textbf{Unknown types}} \\ \cline{3-8}
                                  &                                   & \textbf{Macro F1 (\%)} & \textbf{Accuracy (\%)} & \textbf{Packets (\#)} & \textbf{Time (s)} & \textbf{FPR (\%)} & \textbf{TPR (\%)} \\ \hline
\multirow{3}{*}{UTMobileNet} 
                             & {\color{blue}\textbf{FastFlow}}     & {\color{blue}85.42}  & {\color{blue}86.32} & {\color{blue}12.92 $\pm$ 9.48}   & {\color{blue}0.57 $\pm$ 1.61} & {\color{blue}4.56} & {\color{blue}86.94}  \\ 
                             & Packet seq.                & 79.08  & 79.81 & 8.71 $\pm$ 4.05    & \textbf{0.51 $\pm$ 0.43} & \textbf{3.63} & \textbf{88.24}  \\
                             & Time-int. seq.             & \textbf{87.54}  & \textbf{88.61} & 26.30 $\pm$ 9.10   & 1.63 $\pm$ 0.11 & 4.68 & 83.21  \\  \hline
\multirow{2}{*}{VNAT} 
                            & {\color{blue}\textbf{FastFlow}}      & {\color{blue}96.24}  & {\color{blue}98.03} & {\color{blue}4.16 $\pm$ 1.34}     & {\color{blue}0.033 $\pm$ 1.43} & {\color{blue}\textbf{0}} & {\color{blue}97.32}    \\ 
                            & Packet seq.                 & 94.29  & 96.49 & 3.97 $\pm$ 1.73     & \textbf{0.006 $\pm$ 1.18} & \textbf{0} & \textbf{99.24}    \\
                            & Time-int. seq.              & \textbf{99.40}  & \textbf{99.61} & 7.93 $\pm$ 5.06     & 0.093 $\pm$ 1.32 & \textbf{0} & 98.78    \\ \hline
\multirow{2}{*}{UNIBS} 
                            & {\color{blue}\textbf{FastFlow}}      & {\color{blue}92.30}  & {\color{blue}95.21} & {\color{blue}9.92 $\pm$ 3.92}      & {\color{blue}0.36 $\pm$ 0.82} & {\color{blue}2.54}    & {\color{blue}\textbf{98.16}}  \\ 
                            & Packet seq.                 & 90.51  & 92.37 & 5.44 $\pm$ 3.29      & \textbf{0.27 $\pm$ 1.20} & \textbf{0}       & 97.92  \\
                            & Time-int. seq.              & \textbf{92.60}   & \textbf{96.07} & 18.02 $\pm$ 9.83     & 0.48 $\pm$ 0.42 & 3.18    & 94.4   \\ \hline
\end{tabular}
\end{table}

This assessment shows that the slot representation is the best representation to classify known flows, while the packet representation performs well for unknown flows. In terms of inference time efficiency, the packet representation is the best. By design, \textit{FastFlow} inherits the best of each representation. Its known flow classification performance is inferior, but close to the slot representation. In contrast, its unknown flow identification is close to the packet representation, which performs best. Regarding inference efficiency, \textit{FastFlow} is faster than the time representation but slower than the packet representation. 

\begin{figure*}[t!]
    \centering
    \begin{subfigure}[b]{0.32\linewidth}
        \centering
        \includegraphics[width=\linewidth]{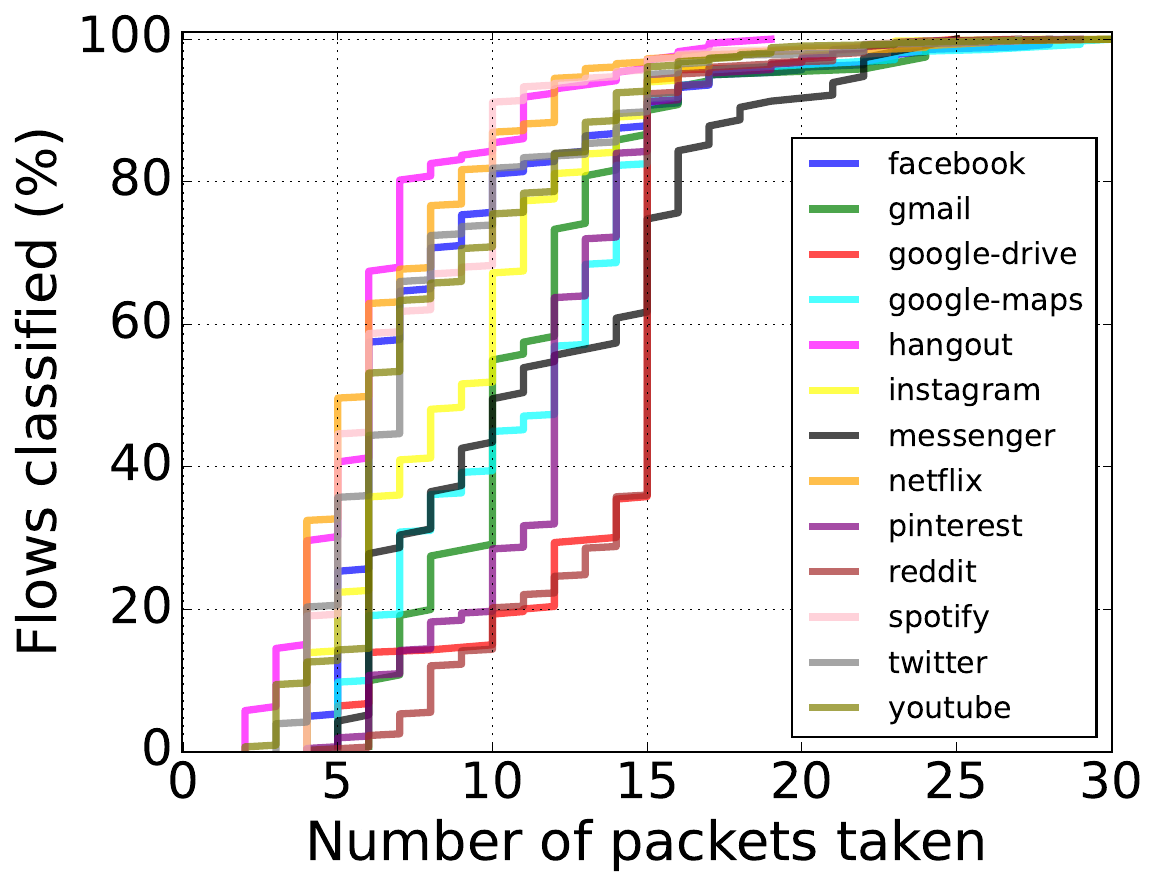}
        \caption{UTMobileNet}
        \label{fig:cdf_utmobilenet}
    \end{subfigure}
    \begin{subfigure}[b]{0.32\linewidth}
        \centering
        \includegraphics[width=\linewidth]{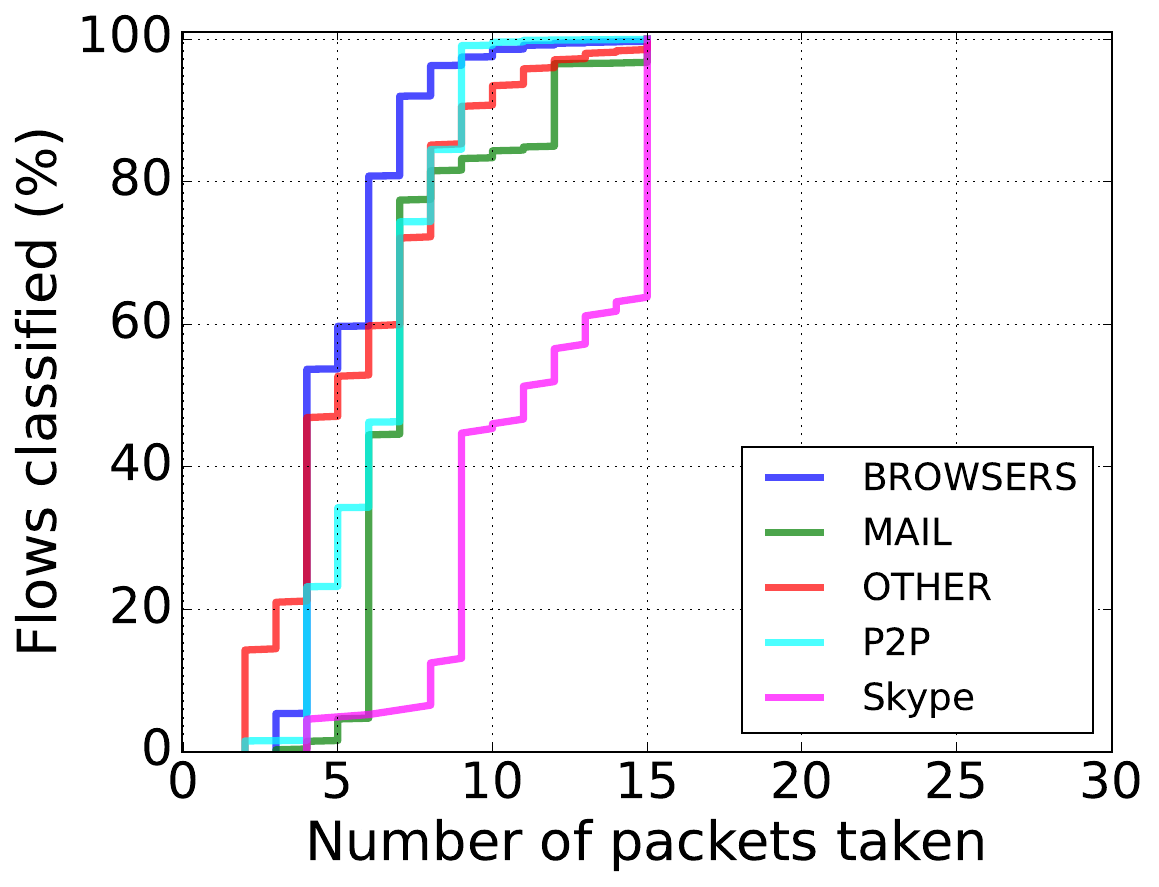}
        \caption{UNIBS}
        \label{fig:cdf_unibs}
    \end{subfigure}
    \begin{subfigure}[b]{0.32\linewidth}
        \centering
        \includegraphics[width=\linewidth]{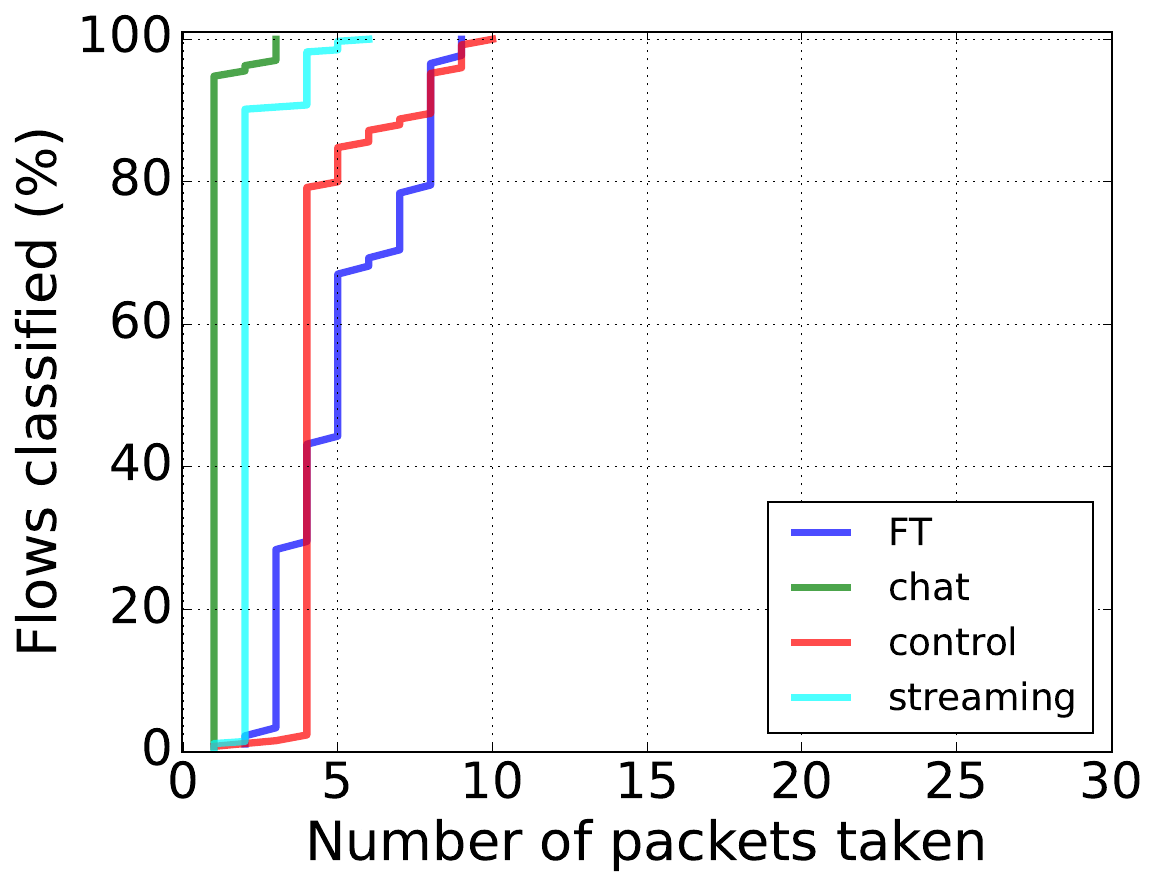}
        \caption{VNAT}
        \label{fig:cdf_vnat}
    \end{subfigure}
    \vspace{-3mm}
    \caption{CDF plots for \textbf{number of packets taken} to classify each flow by \textit{FastFlow} classifiers. }
    \label{fig:cdf_fastflow_packet_plot}
\end{figure*}

\begin{figure*}[t!]
    \centering
    \begin{subfigure}[b]{0.32\linewidth}
        \centering
        \includegraphics[width=\linewidth]{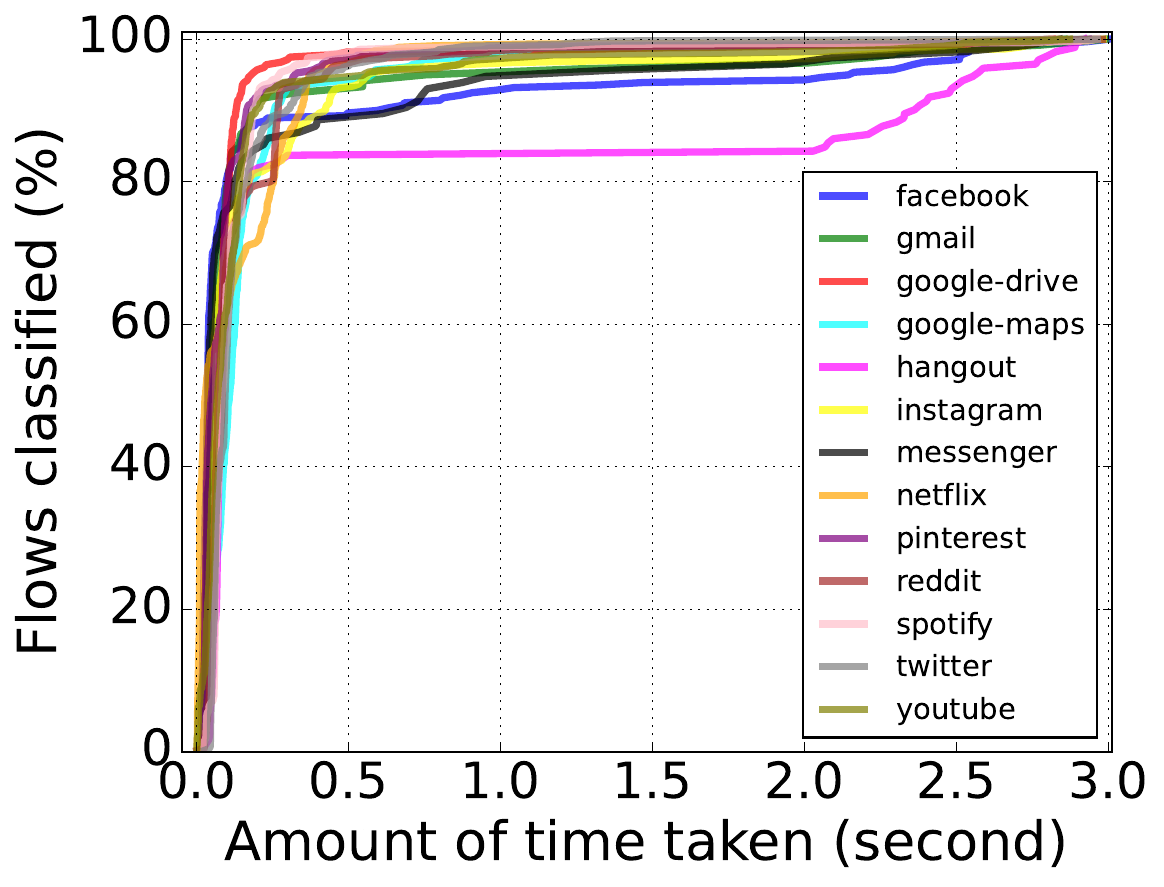}
        \caption{UTMobileNet}
        \label{fig:cdf_utmobilenet}
    \end{subfigure}
    \begin{subfigure}[b]{0.32\linewidth}
        \centering
        \includegraphics[width=\linewidth]{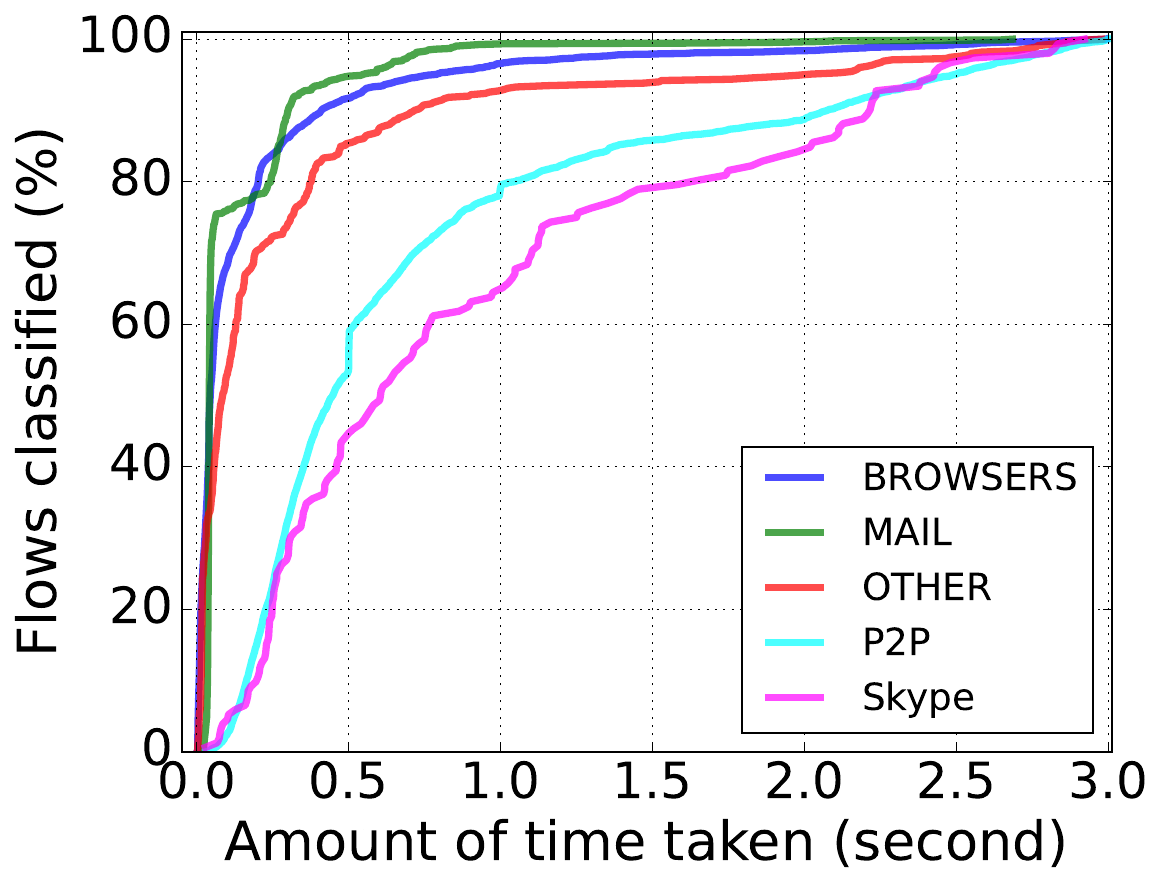}
        \caption{UNIBS}
        \label{fig:cdf_unibs}
    \end{subfigure}
    \begin{subfigure}[b]{0.32\linewidth}
        \centering
        \includegraphics[width=\linewidth]{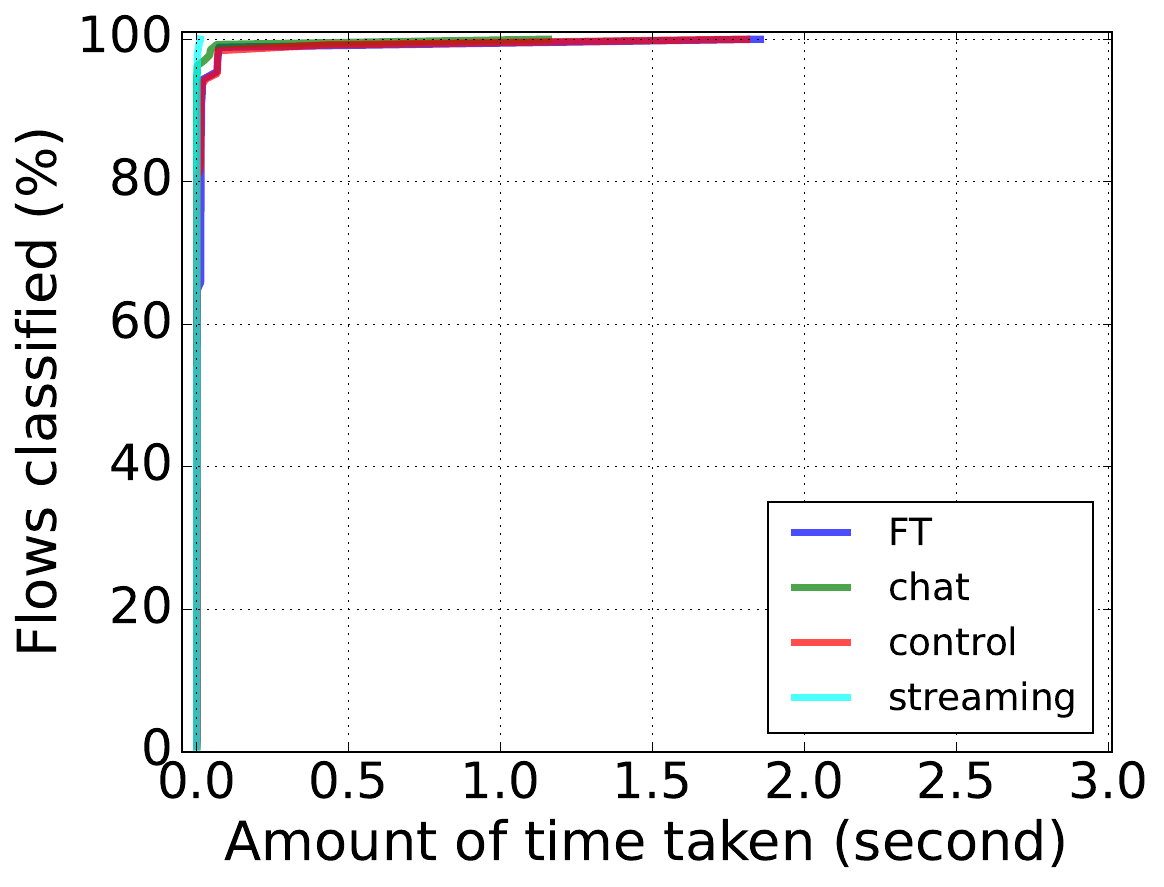}
        \caption{VNAT}
        \label{fig:cdf_vnat}
    \end{subfigure}
    \vspace{-3mm}
    \caption{CDF plots for the \textbf{amount of time taken} to classify each flow by \textit{FastFlow} classifiers. }
    \label{fig:cdf_fastflow_time_plot}
\end{figure*}

Toward this discussion, we now focus on the inference efficiency of \textit{FastFlow}. The results in Tables~\ref{tab:results_table_ood_packet_drop} and~\ref{tab:ablation_results_table_ood_packet_drop} indicate that \textit{FastFlow} is efficient. However, we only report the mean number of packets and inference time for all classes. Therefore,  in Figs.~\ref{fig:cdf_fastflow_packet_plot} and~\ref{fig:cdf_fastflow_time_plot}, we provide details on the inference efficiency of \textit{FastFlow} method for each flow type. Both plots show a \textit{cumulative distribution function} (CDF). Fig.~\ref{fig:cdf_fastflow_packet_plot} shows the fraction of packets per flow type classified by \textit{FastFlow} for a given number of packets. Fig.~\ref{fig:cdf_fastflow_time_plot} similarly shows the fraction of packets per flow classified for a given time threshold in seconds.

Both plots show a high amount of variability across flow types and datasets. Regarding the datasets, UTMobileNet required the largest number of packets (\ie 30) to classify all packets, while UNIBS required half of this amount (\ie 15), and VNAT only needed 10 packets. We reckon that the number of packets may be correlated with the number of flow types in each dataset, as a larger number of flow types makes the recognition of individual flows more challenging. 

Regarding recognizing individual flow types in each dataset, we notice that some can be readily classified, while others require more packets. This contributes to the standard deviation numbers observed in Tables~\ref{tab:results_table_ood_packet_drop} and~\ref{tab:ablation_results_table_ood_packet_drop}. In both cases, these plots indicate the adaptability of \textit{FastFlow} to recognize which flow types are more straightforward to classify.

\subsection{\textit{FastFlow} Deployment in a Live Network}\label{sec:campusDeployment}

Our final evaluation assesses the classification performance and inference efficiency of \textit{FastFlow} in a realistic network deployment with a massive volume of data and a large number of unknown flows. During a one-week deployment from 1st to 7th November 2024, a copy of the traffic exchanged between our university campus border router and the Internet was streamed to our server running \textit{FastFlow} prototype via two 10 Gbps network interfaces for inbound and outbound traffic\footnote{We have obtained ethical clearance approval as detailed in Appendix~\ref{sec:Appendix_Ethics}, which allows us to analyze campus traffic for Internet application classifications without collecting information that can identify users such as university IDs and names.}, respectively. 
The server is configured with an 8-core Intel Xeon E5-2620 CPU and 64GB DDR4 RAM. As depicted in Fig.~\ref{fig:overview}, two LSTM-based models are deployed to analyze real-time packet and slot time-series data, respectively. The number of model executions is directly determined by the number of packets and time-interval slots required to classify a candidate flow, as will be reported in the `Packet (\#)' and `Time (s)' columns of Table~\ref{tab:results_table_deployment} and \ref{tab:deployment_combined_results_table}.
Given the simplified design of the classifier structure, the small dimension of the inference linear layer, and our engineering optimization including huge memory pages, large batch processing sizes, and concurrency, the prototype can be run on the server for our campus deployment (for a peak number of concurrent flows of around 50K and an average of 8.37 packets per flow) without performance issues being observed.
To train our classifiers, we have collected flows labeled by a commercial network traffic classification system from Canopus Networks Pty Ltd one week before the deployment, indicated by the `\# train. flow' column in Table~\ref{table:deployment_dataset_splits}. With the about 1 million labeled flows, our classifiers are trained by a single NVIDIA-4090 GPU. The training process took about 6 hours.

Shown as the `\# live flow' column, during the one-week live deployment, our prototype processed 22.9 million flows that belong to seven application types served by 32 content providers labeled by the commercial system deployed in parallel to our \textit{FastFlow} prototype.

\begin{wraptable}{l}{7cm}
\vspace{-4mm}
    \centering
        \fontsize{7}{9}\selectfont
        \caption{Summary of the deployment data with the number of flows per application and content provider.}
        \vspace{-3mm}
    \begin{tabular}{|c|l|l|l|}
        \hline
        \rowcolor[rgb]{ .906,  .902,  .902} \textbf{Application} & \textbf{Provider} & \textbf{\#live flow} & \textbf{\#train. flow}\\
        \hline
        \multirow{5}{*}{Video Streaming} & MS Stream & 1835585 & 136358 \\
                               & Youtube & 1835508 & 124088 \\
                               & QQ & 592163 & 12424 \\
                                & WeChat & 387700 & 31024 \\
                                & Fastly & 96750 & 15339\\
        \hline
        \multirow{4}{*}{Software Update} & Adobe & 1814389 & 101212 \\
                                        & Windows & 310918 & 34429 \\
                                        &  Apple  &  820493 & 31090 \\
                                        &  Ubuntu  &  10921 & 1763 \\
        \hline
        \multirow{7}{*}{Conferencing} & Discord & 337876 & 38945 \\
                                        & WhatsApp & 30959 & 2589 \\
                                        &  GoogleMeet  &  30034 & 1415 \\
                                        &  MS Teams  &  154832 & 2010 \\
                                        &  Facetime  &  9866 & 815 \\
                                        &  Zoom  &  5424 & 361 \\

        \hline

        \multirow{6}{*}{Social Media} & TikTok & 1492556 & 115313 \\
                                        & Instagram & 1344128 & 73665 \\
                                        &  Facebook  &  759719  & 42566\\
                                        &  LinkedIn  &  185968 & 16363 \\
                                        &  Reddit  &  154804 & 10566 \\
                                        &  Twitter  &  126376 & 6679 \\
                                        
        \hline

        \multirow{5}{*}{File Storage } & Apple iCloud & 908163 & 13808 \\
                                        & MS Sharepoint & 679058 & 25600 \\
                                        &  Dropbox  &  112415 & 11778 \\
                                        &  Google Drive  &  253460 & 8327 \\
                                        &  OneDrive  &  59111 & 3485 \\
                                        
        \hline
        \multirow{4}{*}{Download } 
                                        & AmazonAWS & 1212689 & 35159 \\
                                        & GoogleServices & 882736 & 37417 \\
                                        &  Google  &  755756 & 21928 \\
                                        &  MS DotNET  &  10880 & 13701 \\          
        \hline

       \multirow{2}{*}{Mail} & Microsoft & 738633 & 26950 \\
                             & Google & 87650 & 4216  \\          
        \hline
        {\color{gray}Unknown} & {\color{gray}--}  & {\color{gray}4075187} & {\color{gray}--} \\ \hline
    \end{tabular}
    \label{table:deployment_dataset_splits}
   \vspace{-3mm}
\end{wraptable}

We analyze the performance of \textit{FastFlow} for application types and their content providers. Table~\ref{tab:results_table_deployment} shows the results for the application type. On average, \textit{FastFlow} achieves an accuracy of 91.45\%, and Macro F1 of 90.23\%. The average number of packets is 8.37, and the average time to classify a flow is 0.5 seconds.

Among the seven \textbf{application types}, \textit{FastFlow} achieves very high classification accuracy for conferencing (over 99\%), mail (over 95\%), and software update (over 93\%), all of which have quite deterministic initial packet patterns compared to other application types. For mail and software, such patterns become confidently clear within the first eight packets with slight variations within 3 or 4 packets. In comparison, conferencing flows require about 3 to 18 packets for confident classification by \textit{FastFlow} due to the diversified flow functionalities such as for video, audio, chats, and screen sharing as discussed in \cite{zoom_IMC_passive_2022}, each can have a unique initial packet sequence for function-specific requests.

The other four application types have decent classification performance with less than 10 packets and 1 to 2 seconds for over 85\% accuracy or macro F1 compared to existing literature. The exception is social media flows, which are only classified with about 82\% accuracy using up to 20 packets in 4 seconds. As we can see in the provider list of the social media application type in Table~\ref{table:deployment_dataset_splits}, the flows labeled as social media by the commercial system belong to providers offering a mixed set of services such as short videos (\eg Tiktok), picture sharing (\eg Instagram), online social platforms (\eg Facebook) and forums (\eg Reddit), which are inherently different in the content delivered via network flows compared to each other. With a maximum time step threshold $C_{unk}$ value tuned as 25 packets, our classifiers can accurately detect 91.94\% unknown flows that are labeled by the commercial system. The remaining unknown flows are classified as mostly `social media' and `conferencing', followed by `download', as reported in the `Unknown FPR' column of Table~\ref{tab:results_table_deployment}. We acknowledge that those false positives may not be truly misclassification as we use the labels provided by the commercial system as an estimation of ground truths, which has no means of always being correct, particularly for the applications (\eg social media and software update) that have providers not included in the ground-truth labels.

We also developed and deployed flow classifiers to determine the \textbf{content provider} of each flow with its identified application type. Given that classifying content providers within each application type reduces the classification scope and possible variations within the same flow type, we observe better performance for both accuracy and speed in all subsequent content provider classifiers compared to its preliminary application type classifiers.
We show the classification performance of our content provider classifiers for two representative application types, including video streaming and software update in Table~\ref{tab:deployment_combined_results_table}.

For video streaming content providers, our classifiers achieve superior performance in the three content providers primarily providing streaming services, including Microsoft Stream (used by our university for organizational video content such as lecture recording), YouTube, and Fastly. Over 97\% accuracy is achieved with about 3 to 4 initial packets (less than 0.15 seconds) per flow. The video flows supported by QQ and WeChat, mainly known as social media applications with video streaming as their side features, require an average of 5 to 8 packets (less than 0.2 seconds) for classification accuracy over 89\%. This shows the complexity of flow patterns for content providers that offer services across application types. Notably, all flows labeled (by the commercial system) as video flows get confidently classified by their provider types instead of being classified as the `unknown' type shown in the last column of Table~\ref{tab:deployment_combined_results_table}.

Software update has all its popular content providers accurately (over 93\%) predicted with about six initial packets (less than 0.1 seconds) per flow. Windows and Apple, which have a diversified firmware catalog potentially supported by different product teams, are classified with lower accuracies (94.20\% and 93.12\%) compared to Adobe and Ubuntu (98.24\% and 99.38\%), which are with unified firmware/software portals for updates. From the last column of Table~\ref{tab:deployment_combined_results_table}, a minority (less than 0.48\%) of Adobe and Windows flows labeled by the commercial system are misclassified as `unknown' by our \textit{FastFlow} classifier, suggesting that further improvements on the fine-grained labeled dataset are needed to train a more accurate classifier on those application/provider types.

\begin{table}[t!]
    \centering
        \fontsize{8}{10}\selectfont
        \caption{Performance metrics of flow classification by \textit{FastFlow} for application type.}
        \vspace{-3mm}
    \begin{tabular}{|c|c|c|c|c|c|}
        \hline
       \rowcolor[rgb]{ .906,  .902,  .902} \textbf{Application} & \textbf{Macro F1 (\%)} & \textbf{Accuracy (\%)} & \textbf{Packets (\#)} & \textbf{Time (s)} & \textbf{Unknown FPR (\%)} \\
        \hline
        Video Streaming & 89.90 & 88.73 & 8.31 $\pm$ 4.38 & 0.21 $\pm$ 1.32 & 0.00 \\
        Software Update & 92.50 & 93.18 & 7.22 $\pm$ 4.50 & 0.10 $\pm$ 1.06 & 0.00 \\
        Conferencing &  98.63 & 99.82 & 10.57 $\pm$ 7.27 & 1.34 $\pm$ 3.89 & 4.34 \\
        Social Media &  81.91 & 82.42 & 11.89 $\pm$ 5.65 & 1.59 $\pm$ 2.38 & 5.17 \\
        File Storage  & 89.13 & 84.75 & 8.83 $\pm$ 4.11 & 0.23 $\pm$ 2.17 & 0.00 \\
        Download  & 83.60 & 87.97 & 10.10 $\pm$ 6.22 & 0.20 $\pm$ 1.21 & 1.43 \\
        Mail  & 93.66 & 95.20 & 6.12 $\pm$ 3.33 & 1.68 $\pm$ 9.58 & 0.00 \\
        \textbf{Unknown} & -- & 91.94 & 20 $\pm$ 0.00 & 0.77 $\pm$ 2.69 & -- \\
        \textbf{\color{blue}Average} & {\color{blue}90.23} & {\color{blue}91.45} & {\color{blue}8.37} & {\color{blue}0.5} & {\color{blue}--} \\
        \hline
    \end{tabular}
    \label{tab:results_table_deployment}
    \vspace{-3mm}
\end{table}

As discussed earlier for the result of application types, social media flows are inherently diversified in their functionalities, which also leads to a mediocre classification performance (80\% to 90\% in accuracy and Marco F1) to classify the content provider of each flow. This necessitates a ground-truth dataset with well-engineered flow types within the application type for training purposes, which is not within the scope of this paper.
Similar insights are observed for other application types and are not explicitly discussed.

\textbf{System deployment considerations:} Here we discuss two practical considerations when deploying \textit{FastFlow} in a live network.
First, according to our performance benchmarking, around 30K concurrent flows can be processed by a time-series classifier on a single Intel Xeon E5-2620 CPU core of the server. Therefore, our 8-core CPU is sufficient for processing our live campus traffic with up to 50K concurrent flows. While not used in our current deployment, we have also benchmarked on a NVIDIA-4090 GPU that the two classifiers in \textit{FastFlow} can process up to 200K concurrent flows. For larger workloads, we suggest two \textbf{scaling-up options}, including leveraging GPU servers that are more efficient in executing neural network models and setting up parallel computing nodes, each processing a subnet of the monitored network.
Second, the classification performance of machine learning classifiers can be significantly impacted by the \textbf{quality of the training data}. For \textit{FastFlow}, its performance for both classification of known flow types and unknown flow detection is directly determined by the quality of the flow data used in the training process. In our proof-of-concept research prototype, we used a commercial network traffic classification system for flow labels. In industrial practice, such localized training data are often obtained from ISP digital twins or by the service team of a network observability platform.

\begin{table}[t!]
    \centering
    \fontsize{7.5}{9}\selectfont
    \caption{Performance of flow classification by \textit{FastFlow} for representative application providers.}
    \vspace{-3mm}
    \begin{tabular}{|c|c|c|c|c|c|c|}
        \hline
       \rowcolor[rgb]{ .906,  .902,  .902} \textbf{Application} & \textbf{Provider} & \textbf{Macro F1 (\%)} & \textbf{Accuracy (\%)} & \textbf{Packet (\#)} & \textbf{Time (s)} & \textbf{Unknown FPR (\%)} \\
        \hline
        \multirow{5}{*}{Video Streaming} 
         & Microsoft & 98.70 & 99.29 & 3.25 $\pm$ 2.32 & 0.05 $\pm$ 0.44 & 0.00 \\
          & YouTube & 97.98 & 97.14 & 4.25 $\pm$ 3.98 & 0.15 $\pm$ 1.51 & 0.00 \\
           & QQ & 89.77 & 86.32 & 8.32 $\pm$ 4.68 & 0.19 $\pm$ 0.36 & 0.00 \\
            & WeChat & 91.01 & 91.56 & 5.78 $\pm$ 5.43 & 0.11 $\pm$ 0.40 & 0.00 \\
        & Fastly & 99.05 & 99.33 & 4.76 $\pm$ 1.56 & 0.01 $\pm$ 0.02 & 0.00 \\

        \hline
        \multirow{4}{*}{Software Update} & Adobe & 98.09 & 98.24 & 4.75 $\pm$ 2.44 & 0.02 $\pm$ 0.09 & 0.48 \\
                                        & Windows & 93.84 & 94.20 & 6.40 $\pm$ 2.97 & 0.08 $\pm$ 0.51 & 0.38 \\
                                        & Apple & 95.20 & 93.12 & 6.61 $\pm$ 4.16 & 0.06 $\pm$ 0.46 & 0.00 \\
                                        & Ubuntu & 98.77 & 99.38 & 6.63 $\pm$ 1.55 & 0.13 $\pm$ 0.15 & 0.00 \\
                                      
        \hline
         \multirow{4}{*}{Conferencing} & Discord & 98.74 & 99.70 & 1.20 $\pm$ 1.59 & 0.04 $\pm$ 0.03 & 0.00 \\
                                       
                                         & Whatsapp & 99.22 & 99.38 & 2.99 $\pm$ 2.30  & 0.39 $\pm$ 2.45 & 0.00 \\
                                        & GoogleMeet & 98.41 & 96.87 & 4.00 $\pm$ 4.09 & 0.13 $\pm$ 0.05 & 0.03 \\
                                         &MS Teams & 98.68 & 99.20 & 2.51 $\pm$ 1.45 & 0.57 $\pm$ 2.51 & 0.00 \\
                                        & FaceTime & 97.77 & 95.65 & 3.65 $\pm$ 3.60 & 0.38 $\pm$ 1.80 & 0.00 \\
                                        & Zoom & 97.41 & 98.62 & 4.12 $\pm$ 4.45 & 0.99 $\pm$ 3.62 & 0.07 \\

        \hline
        \multirow{4}{*}{Social Media} 
                                         & TikTok & 83.51 & 86.36 & 6.31 $\pm$ 3.52 & 0.13 $\pm$ 0.18 & 0.00  \\
                                        & Instagram & 85.17 & 88.97 & 11.16 $\pm$ 5.57 & 0.29 $\pm$ 1.39 & 0.00 \\
                                         & Facebook & 82.35 & 82.12 & 9.90 $\pm$ 5.23 & 0.06 $\pm$ 0.28 & 0.05 \\
                                        & LinkedIn & 81.16 & 89.09 & 10.52 $\pm$ 5.26 & 0.27 $\pm$ 1.62 & 0.00 \\
                                        & Reddit & 84.61 & 88.85 & 9.09 $\pm$ 4.62 & 0.67 $\pm$ 4.10 & 0.00 \\
                                        & Twitter & 84.06 & 88.14 & 3.85 $\pm$ 4.11 & 0.02 $\pm$ 0.10 & 0.00 \\

        \hline
         \multirow{4}{*}{File Storage} 
          & Apple iCloud & 96.44 & 95.00 & 11.89 $\pm$ 4.62 & 0.05 $\pm$ 0.10 & 4.76 \\
           & MS Sharepoint & 91.94 & 95.65 & 7.35 $\pm$ 4.40 & 0.05 $\pm$ 0.27 & 2.42 \\
            & Dropbox & 96.42 & 97.29 & 8.12 $\pm$ 2.63 &0.08 $\pm$ 0.12 &0.00  \\
         & Google Drive & 97.77 & 96.24 & 3.85 $\pm$ 4.11 & 0.02 $\pm$ 0.10 & 1.48 \\
        & OneDrive & 88.37 & 84.73 & 9.63 $\pm$ 3.95 & 0.03 $\pm$ 0.07 & 0.00 \\

        \hline
    \end{tabular}
    \label{tab:deployment_combined_results_table}
    \vspace{-3mm}
\end{table}

\section{Conclusion}

In this paper, we present \textit{FastFlow}, a time-series early flow classification method that can be practically deployed in large networks such as ISPs at runtime. By developing a dual-grained time-series flow representation scheme and innovating a time-series flow classifier architecture trained with reinforcement learning techniques, \textit{FastFlow} is the first of its kind that addresses three key deployment challenges in large networks, including accurate classification with the estimated minimal number of initial packets in each candidate flow, robust to packet sequence disorders, and capable of detecting unknown flow types. We extensively validate the classification performance of \textit{FastFlow} using public datasets and compare its performance with ablation alternatives and state-of-the-art methods. \textit{FastFlow} is implemented and deployed in a large campus network to classify application types and content providers of network flows. The deployment insights showcase that \textit{FastFlow} classifiers can accurately classify flows for their applications and content providers with only about 10 initial packets of each flow in less than one or two seconds, and are able to detect flows that do not belong to a known type. 

\section*{Acknowledgement}
We thank our shepherd Francesco Bronzino and the anonymous reviewers for their insightful feedback. This work is supported by the Australian Government's Cooperative Research Centres Projects (CRC-P) Grant CRCPXIV000099.

\bibliographystyle{ACM-Reference-Format}
\bibliography{fastFlow}

\appendix

\section{Ethics}\label{sec:Appendix_Ethics}
We have obtained ethical clearance from our university ethics board (UNSW Human Research Ethics Advisory Panel approval number HC211007) which allows us to analyze campus traffic for Internet applications without being able to access user identities such as ID numbers and names. In our campus deployment, insights into application types and providers were reported in an aggregated manner, preserving anonymity rather than identifying specific users. In our analysis, no attempt was made to associate network flows with personal identities.

\section{Additional Lab Evaluation Results of \textit{FastFlow} Classification Performance}\label{sec:Appendix_Eval}

\subsection{Speed of \textit{FastFlow} Classifiers with Only Packet or Slot Data Sequence}\label{sec:appendix_FastFlow_Speed}
\subsubsection{Using Only Packet Data Sequence}

\begin{figure*}[h!]
    \centering
    \begin{subfigure}[b]{0.3\linewidth}
        \centering
        \includegraphics[width=\linewidth]{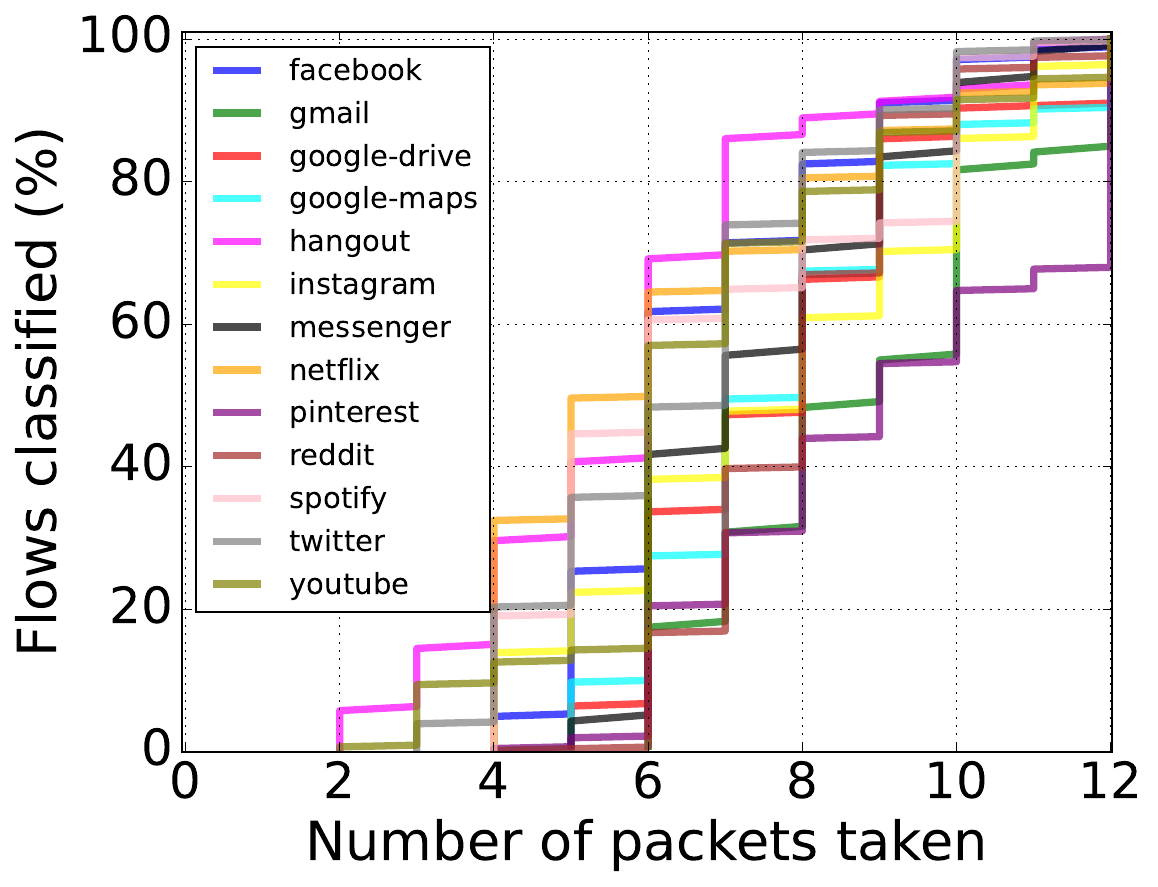}
        \caption{UTMobilenet}
        \label{fig:cdf_packet_utmobilen}
    \end{subfigure}
    \begin{subfigure}[b]{0.3\linewidth}
        \centering
        \includegraphics[width=\linewidth]{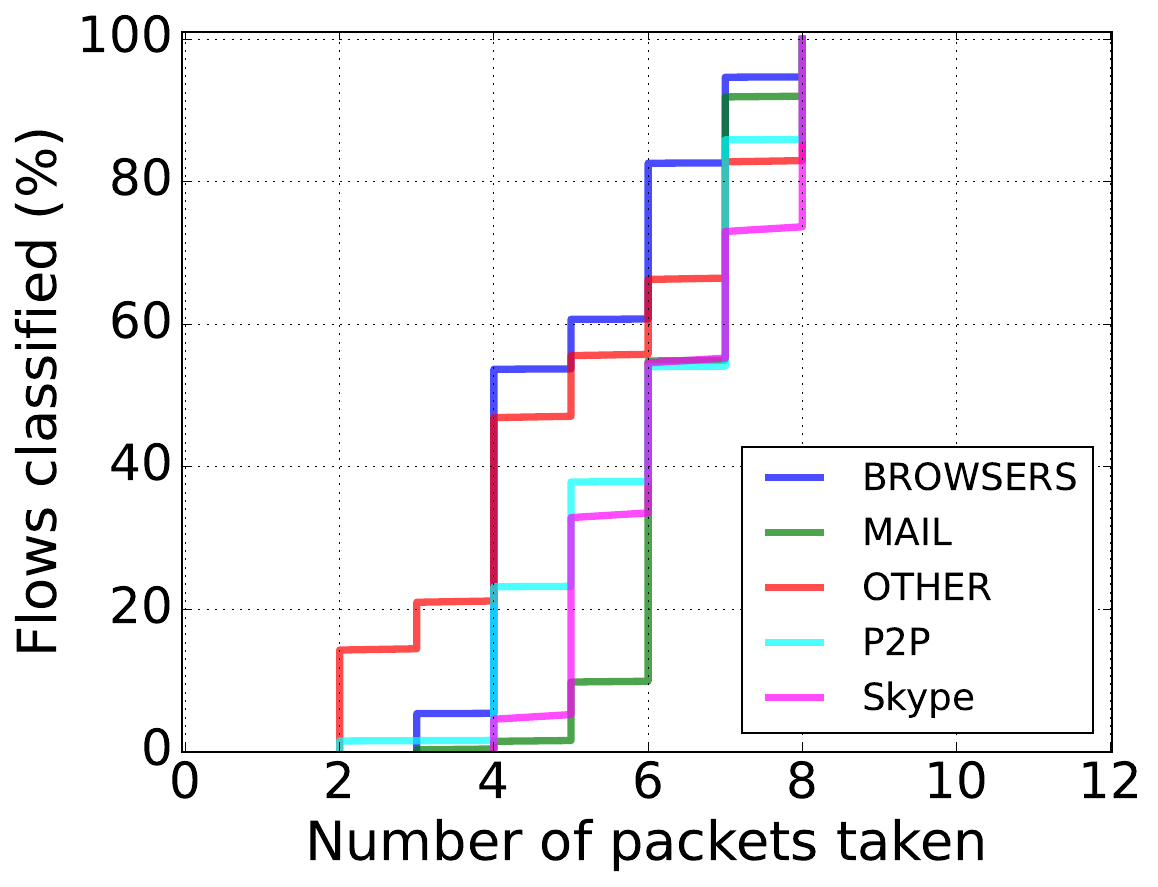}
        \caption{UNIBS}
        \label{fig:cdf_packet_unibs}
    \end{subfigure}
    \begin{subfigure}[b]{0.3\linewidth}
        \centering
        \includegraphics[width=\linewidth]{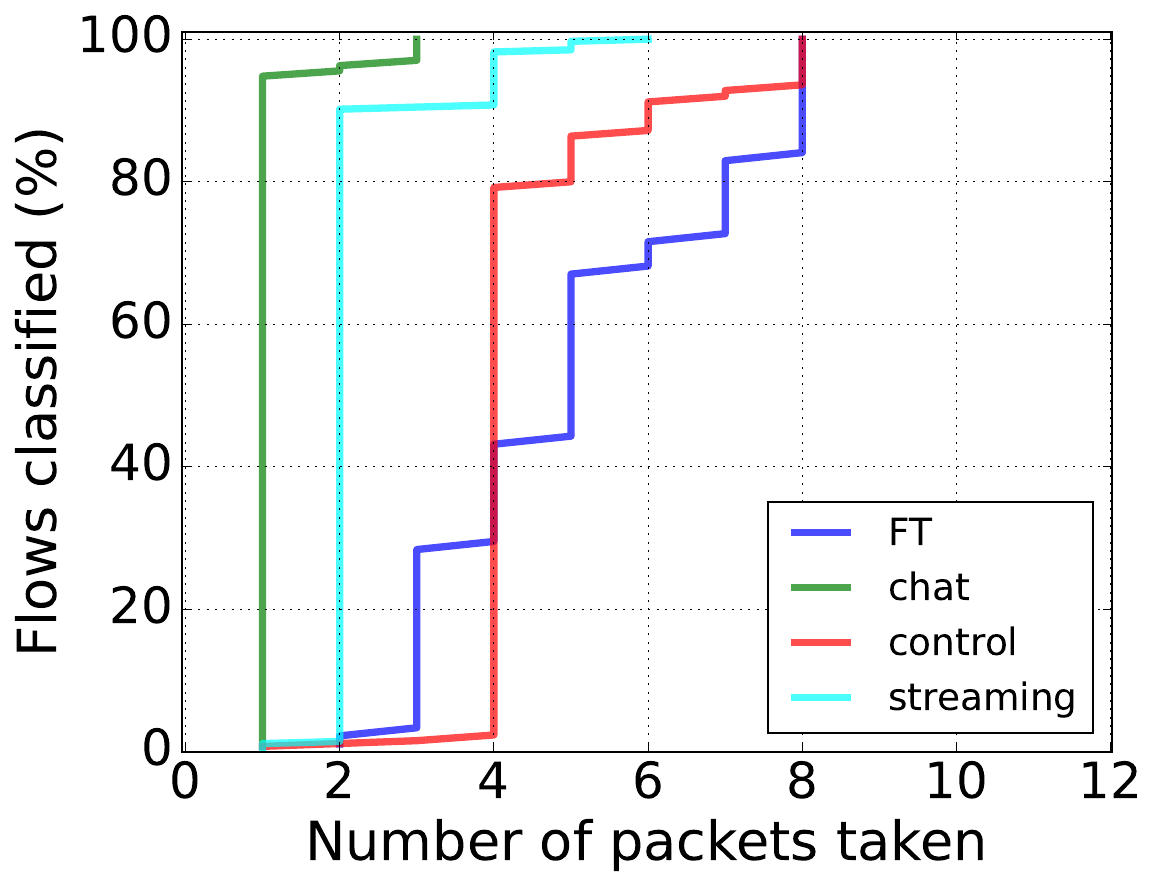}
        \caption{VNAT}
        \label{fig:cdf_packet_vnat}
    \end{subfigure}
    \vspace{-4mm}
    \caption{CDF plots for number of packets taken by \textit{FastFlow} classifiers on \textbf{only packet} data sequence. }
    \label{fig:cdf_packet_plot}
\end{figure*}

Fig.~\ref{fig:cdf_packet_plot} contains CDF plots showing the number of packets taken to classify flows by the classifiers that only consume the packet data sequence. For the three datasets, the number of packets required for flow classification is usually less than 8. However, as discussed before, our classifiers that only use packet data sequence cannot achieve decent performance when there are packet drops and retransmissions in a candidate flow.
 
\begin{figure*}[h!]
    \centering
    \begin{subfigure}[b]{0.3\linewidth}
        \centering
        \includegraphics[width=\linewidth]{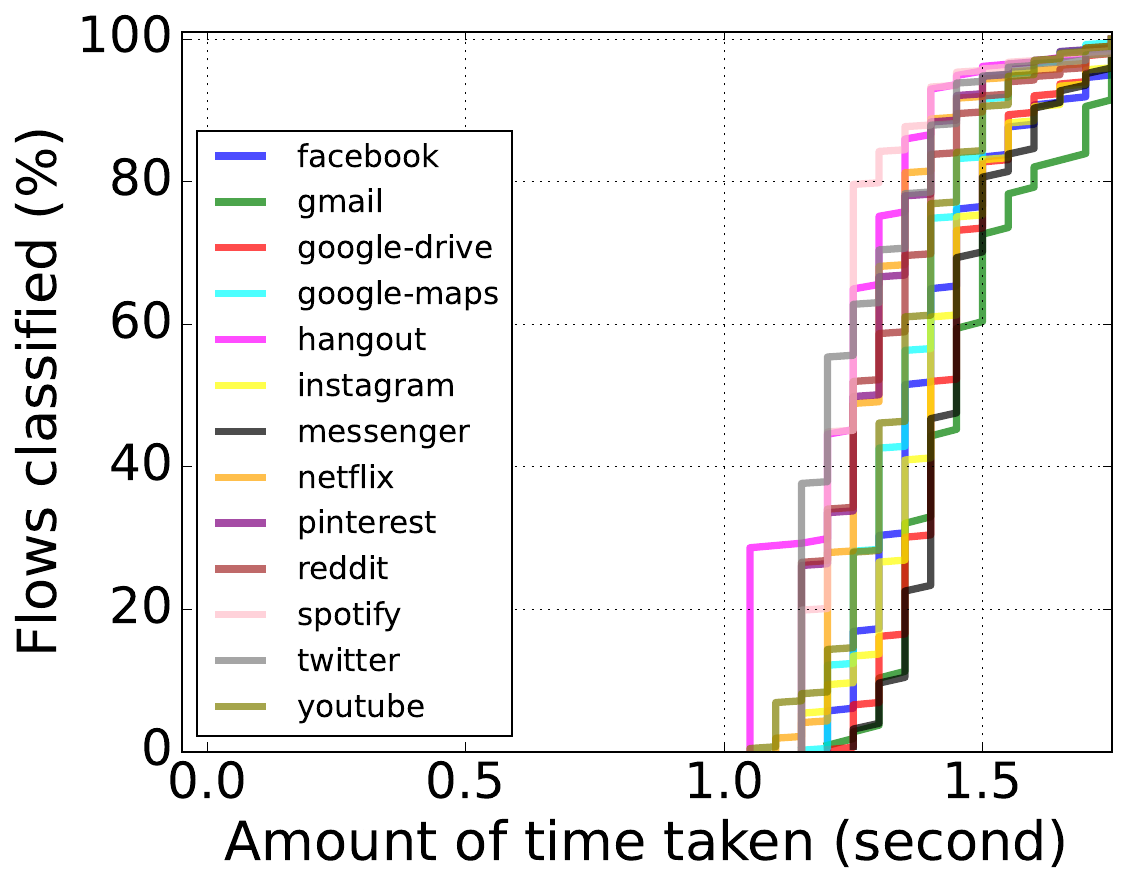}
        \caption{UTMobileNet}
        \label{fig:cdf_time_utmobilenet}
    \end{subfigure}
    \begin{subfigure}[b]{0.3\linewidth}
        \centering
        \includegraphics[width=\linewidth]{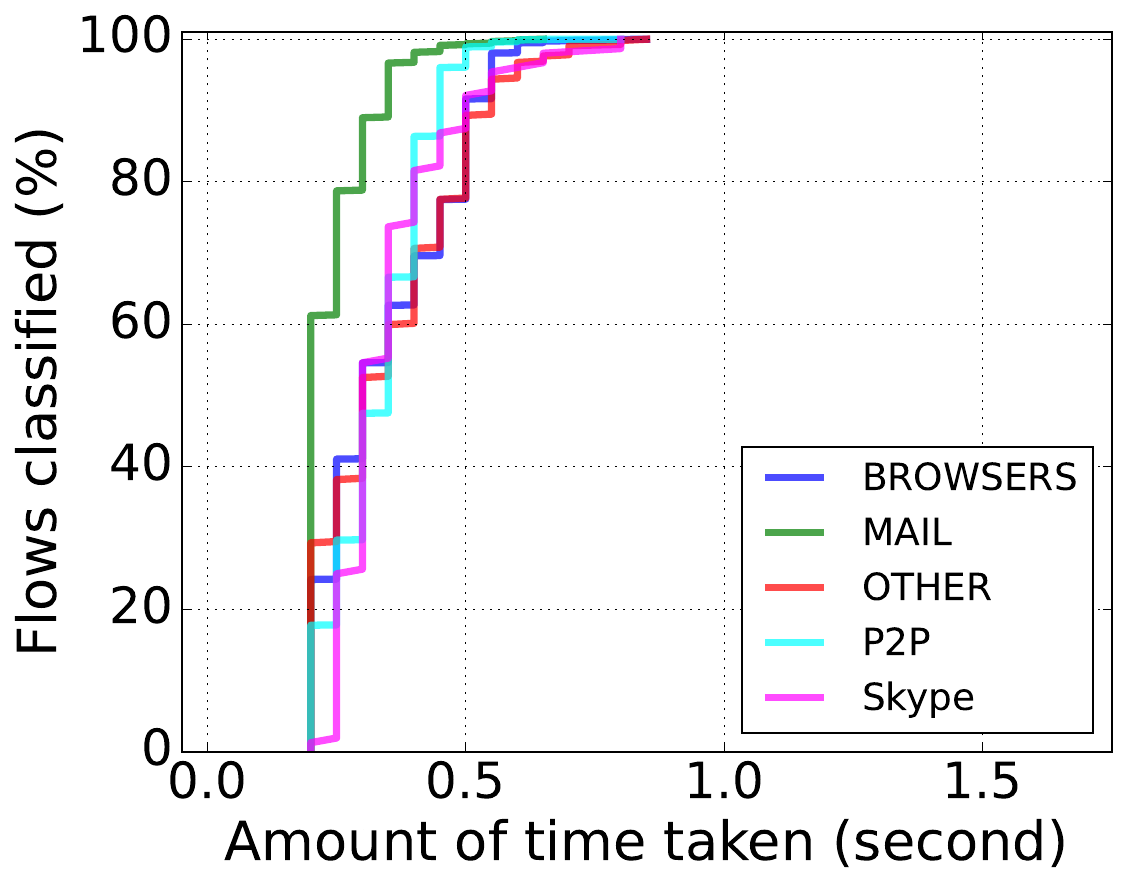}
        \caption{UNIBS}
        \label{fig:cdf_time_unibs}
    \end{subfigure}
    \begin{subfigure}[b]{0.3\linewidth}
        \centering
        \includegraphics[width=\linewidth]{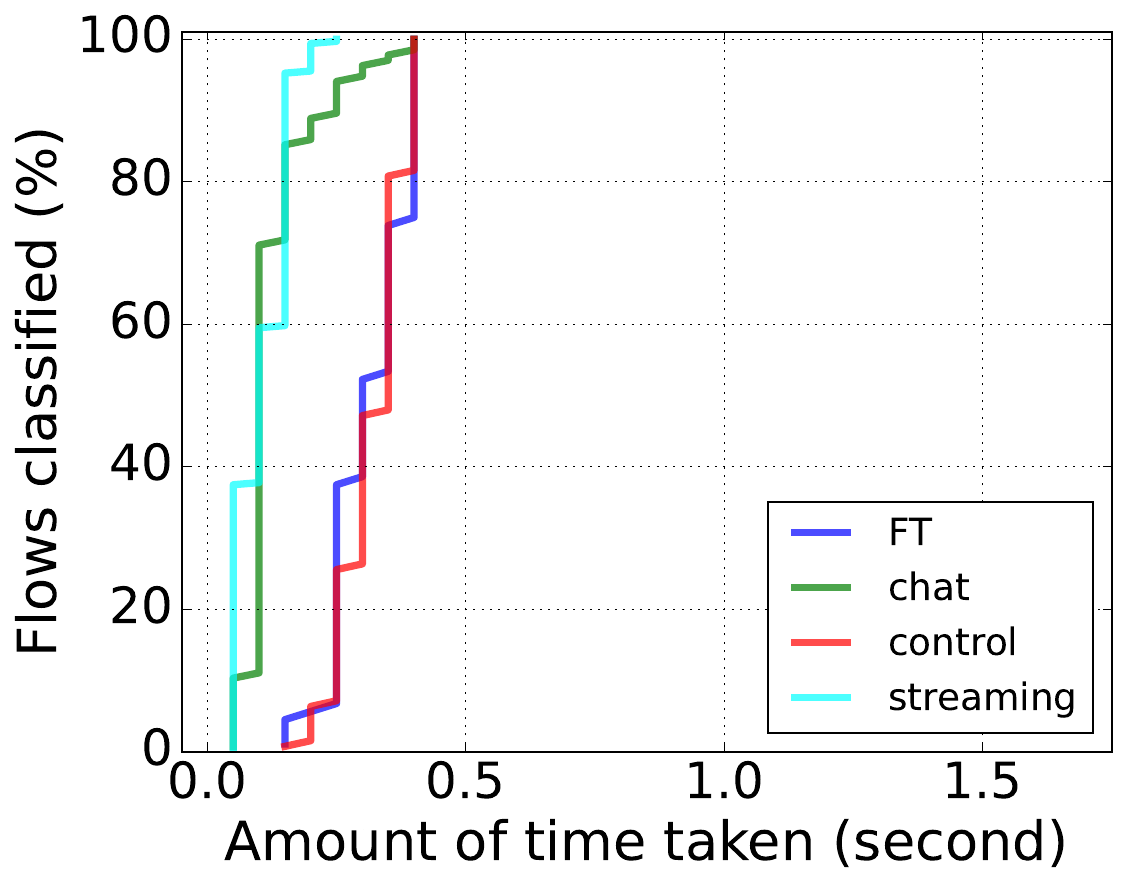}
        \caption{VNAT}
        \label{fig:cdf_time_vnat}
    \end{subfigure}
    \vspace{-3mm}
    \caption{CDF plots for amount of time taken by \textit{FastFlow} classifiers on \textbf{only slot} data sequence.}
    \label{fig:cdf_time_plot}
\end{figure*}

\begin{table}[h!]
\centering
\caption{Classification performance of \textit{FastFlow} on ideal network flow dataset without packet sequence disorder or unknown flow type.}
\vspace{-3mm}
\label{tab:results_table}
\fontsize{7}{9}\selectfont
\begin{tabular}{|l|l|c|c|c|c|}
\hline
\rowcolor[rgb]{ .906,  .902,  .902} \textbf{Dataset} & \textbf{Method} & \textbf{Macro F1 (\%)} & \textbf{Accuracy (\%)} & \textbf{Packets Taken (\#)} & \textbf{Time Taken (s)} \\ \hline
\multirow{6}{*}{UTMobileNet} 
   & \textbf{\color{blue}FastFlow} & {\color{blue}89.44}    & {\color{blue}90.54}     & \color{blue}{13.96  $\pm$ 5.94} & \color{blue}{0.70 $\pm$ 2.45} \\
    & {\color{gray}Packet seq.} & {\color{gray}90.83}    & {\color{gray}92.20}  & {\color{gray}9.07 $\pm$ 3.91} & {\color{gray}0.47 $\pm$ 2.73}    \\
                    
    & {\color{gray}Time-int. seq.}       & {\color{gray}88.74}    & {\color{gray}91.50}    & {\color{gray}26.50 $\pm$ 12.95} & {\color{gray}1.32 $\pm$ 3.80}   \\
    & {\color{purple}GGFast \cite{ggfast}}       & {\color{purple}88.04}     & {\color{purple}90.56}     & {\color{purple}50} & {\color{purple}2.46 $\pm$ 1.98}\\
    & {\color{purple}Grad-BP \cite{lessons_learned_from_commercial}} & {\color{purple}89.08} & {\color{purple}88.11} & {\color{purple}100(TCP) 10(UDP)} & {\color{purple}4.77 $\pm$ 2.31}\\
   & Pkt.-5  & 69.97 & 74.33 & 5 $\pm$ 0.00 & 0.24 $\pm$ 1.95  \\         
                             & Pkt.-45 & 89.76     & 91.06     & 45 $\pm$ 0.00 &  2.30 $\pm$ 1.47\\
                             & Time-int.-5 & 73.41  &   74.02        & 4.91 $\pm$ 2.72  & 0.25 \\ 
                             & Time-int.-45 & 86.47  & 91.10   & 44.92 $\pm$ 7.21  & 2.25 \\          

                             \hline
\multirow{6}{*}{VNAT}      
 & \textbf{\color{blue}FastFlow} & {\color{blue}95.00}  & {\color{blue}96.31}  & {\color{blue}4.36  $\pm$ 6.35} &  {\color{blue}0.02 $\pm$ 1.29}  \\
 & {\color{gray}Packet seq.}    & {\color{gray}95.22}  & {\color{gray}96.60}   & {\color{gray}2.94 $\pm$ 2.31} & {\color{gray}.007 $\pm$ 0.15}  \\
 &  {\color{gray}Time-int. seq.}   & {\color{gray}92.25}  & {\color{gray}95.17}  & {\color{gray}10.57 $\pm$ 4.20} & {\color{gray}0.15 $\pm$ 0.24}   \\
  & {\color{purple}GGFast \cite{ggfast}} & {\color{purple}77.56}  & {\color{purple}90.81} & {\color{purple}50} & {\color{purple}1.07 $\pm$ 1.35}\\
 &{\color{purple}Grad-BP \cite{lessons_learned_from_commercial}} & {\color{purple}93.70} & {\color{purple}95.92} & {\color{purple}100(TCP) 10(UDP)}  & {\color{purple}1.92 $\pm$ 1.35}\\
 & Pkt.-5 & 95.76 & 96.78 & 5 $\pm$ 0.00 & 0.03 $\pm$ 0.62 \\
                             & Pkt.-45  & 87.87  & 87.92 &  45 $\pm$ 0.00 & 0.90 $\pm$ 2.86\\
                             & Time-int.-5 &  90.10 &  92.17 & 14.89 $\pm$ 5.24  & 0.25 $\pm$ 0.00  \\
                             & Time-int.-45 & 80.71 & 82.09 & 107.45 $\pm$ 18.19 & 2.25 $\pm$ 0.00 \\

                             \hline
\multirow{6}{*}{UNIBS}  
 & \textbf{\color{blue}FastFlow} & {\color{blue}95.87}  & {\color{blue}98.34}     & {\color{blue}5.42  $\pm$ 2.19} &   {\color{blue}0.28 $\pm$ 0.60}    \\
 & {\color{gray}Packet seq.}  & {\color{gray}95.87}  & {\color{gray}98.34}  & {\color{gray}5.42 $\pm$ 2.19} & {\color{gray}0.28 $\pm$ 0.60} \\
& {\color{gray}Time-int. seq.}  & {\color{gray}93.57}  & {\color{gray}97.05}  & {\color{gray}10.26 $\pm$ 8.66} & {\color{gray}0.43 $\pm$ 1.68}   \\
 & {\color{purple}GGFast \cite{ggfast}}       & {\color{purple}91.97}     & {\color{purple}95.03}     & {\color{purple}50} & {\color{purple}1.79 $\pm$ 3.40}\\
& {\color{purple}Grad-BP \cite{lessons_learned_from_commercial}} & {\color{purple}93.87}& {\color{purple}97.22} & {\color{purple}100(TCP) 10(UDP)} & {\color{purple}3.28 $\pm$ 2.05} \\
 & Pkt.-5 & 86.68 & 93.82 & 5 $\pm$ 0.00 & 0.26 $\pm$ 0.46 \\
                             & Pkt.-45  & 92.30     & 92.44     & 45 $\pm$ 0.00 & 1.56 $\pm$ 3.06 \\
                             & Time-int.-5 & 72.41 & 76.32 & 4.84 $\pm$ 2.37 & 0.25  \\
                             & Time-int.-45 & 91.60 & 93.39 & 64.60 $\pm$ 9.75 & 2.25  \\

                             \hline
\end{tabular}
\end{table}

\subsubsection{Using Only Time-Interval Slot Data Sequence}

Fig.~\ref{fig:cdf_time_plot} contains CDF plots for the amount of time needed for classifiers that only uses time-interval slot data sequence. Compared to the CDF plots in Fig.~\ref{fig:cdf_fastflow_time_plot} for \textit{FastFlow} classifiers using both packet and slot data sequences, we can conclude that \textit{FastFlow} performs significantly faster than the classifiers that only use slot data sequence.

\subsection{Flow Classification Performance}
\subsubsection{Ideal Conditions without Packet Sequence Disorder and Unknown Flow Type} 

Table~\ref{tab:results_table} shows the classification results without augmenting the three public datasets. As expected, the macro-F1 and accuracy scores for all methods are higher compared to those with packet sequence disorder introduced to the datasets. \textit{FastFlow} performs slightly better than two state-of-the-art methods (\ie \textit{GGFast} and \textit{Grad-BP}) on the three datasets. Also, \textit{FastFlow} requires much smaller number of packets on average to classify flows compared to the state-of-the-art methods.
For the ablation studies, on the datasets without introduced packet sequence disorder, \textit{FastFlow} classifiers that only use packet data (\ie `Packet seq.') achieved equivalent accuracies on the three datasets compared to the classifiers that use both packets and time-interval slots. Classifiers that only use time-interval slot data (\ie `Time-int. seq.') are less accurate in such scenarios.
\textit{FastFlow} also outperforms all classifiers using fixed length of input data.

\subsubsection{With Packet Sequence Disorders but not Unknown Flow Type}
\begin{table}[t!]
\centering
\caption{Classification performance of \textit{FastFlow} with packet sequence disorders in each candidate flow.}
\label{tab:results_table_packet_drop}
\fontsize{7}{9}\selectfont
\begin{tabular}{|l|l|c|c|c|c|}
\hline
\rowcolor[rgb]{ .906,  .902,  .902} \textbf{Dataset} & \textbf{Method} & \textbf{Macro F1 (\%)} & \textbf{Accuracy (\%)} & \textbf{Packets Taken (\#)} & \textbf{Time Taken (s)} \\ \hline
\multirow{9}{*}{UTMobileNet} 
                    & {\color{blue}\textbf{FastFlow}}     & {\color{blue}85.32}    & {\color{blue}87.21}     & {\color{blue}12.83 $\pm$ 6.74} & {\color{blue}0.93 $\pm$ 1.67}  \\

                    & {\color{gray}Packet seq.} & {\color{gray}80.30}      &{\color{gray}82.05} & {\color{gray}9.29 $\pm$ 4.00} & {\color{gray}0.54 $\pm$ 0.76}  \\    
                            
                    & {\color{gray}Time-int. seq.}  & {\color{gray}86.69}     & {\color{gray}88.20}   & {\color{gray}18.21 $\pm$ 9.96} & {\color{gray}1.37 $\pm$ 2.36}   \\ 

                    & {\color{purple}GGFast}       & {\color{purple}72.43}    & {\color{purple}78.49}     & {\color{purple}50} & {\color{purple}2.49 $\pm$ 1.61} \\
                    
                    & {\color{purple}Grad-BP} & {\color{purple}82.59} & {\color{purple}84.25} & {\color{purple}100(TCP) 10(UDP)} & {\color{purple}5.02 $\pm$ 1.85} \\

                             & Pkt.-5      & 64.72     & 71.73     & 5 $\pm$ 0.00 & 0.28 $\pm$ .202 \\
                             & Pkt.-45     &81.20     &82.84     & 45 $\pm$ 0.00 & 2.35 $\pm$ 1.94 \\
                             & Time-int.-5 &   70.28 & 72.62 &  3.92 $\pm$ 2.78 & .25 \\
                             & Time-int.-45 & 85.90 & 89.04 & 44.09 $\pm$ 8.53 & 2.25  \\

                             \hline
\multirow{9}{*}{VNAT}        
& {\color{blue}\textbf{FastFlow}}     & {\color{blue}91.32}    & {\color{blue}94.82}   & {\color{blue}4.88 $\pm$ 6.79} & {\color{blue}.061 $\pm$ 0.43} \\
  & {\color{gray}Packet seq.}       & {\color{gray}89.42}    & {\color{gray}93.71}   & {\color{gray}3.03 $\pm$ 2.59}  & {\color{gray}.021 $\pm$ 0.33}   \\
& {\color{gray}Time-int. seq.}         & {\color{gray}91.82}    & {\color{gray}94.89}   & {\color{gray}10.08 $\pm$ 7.2}  & {\color{gray}0.19 $\pm$ 0.24}  \\ 
  & {\color{purple}GGFast}       & {\color{purple}65.22}     & {\color{purple}76.48}     & {\color{purple}50} & {\color{purple}1.28 $\pm$ .641} \\
& {\color{purple}Grad-BP} & {\color{purple}88.62} & {\color{purple}90.33} & {\color{purple}100(TCP) 10(UDP)} & {\color{purple}2.47 $\pm$ 0.94} \\
& Pkt.-5 & 88.91 & 92.34 & 5 $\pm$ 0.00 & 0.04 $\pm$ .399   \\
& Pkt.-45      & 82.10    & 84.20    & 45 $\pm$ 0.00 & 1.11 $\pm$ 2.94 \\
& Time-int.-5 & 89.46 & 91.89 & 12.23 $\pm$ 6.02 & .25  \\
& Time-int.-45 &80.27 & 82.11 & 86.77 $\pm$ 17.02 & 2.25 \\

                            \hline
\multirow{9}{*}{UNIBS}  

 & {\color{blue}\textbf{FastFlow}}     & {\color{blue}92.04}    & {\color{blue}97.28}      & {\color{blue}6.99 $\pm$ 4.17} & {\color{blue}0.40 $\pm$ 0.91} \\
                             
 & {\color{gray}Packet seq.}       & {\color{gray}89.06}      & {\color{gray}94.86}  & {\color{gray}5.59 $\pm$ 2.38}  & {\color{gray}0.39 $\pm$ 2.27} \\
& {\color{gray}Time-int. seq.}         & {\color{gray}92.33}     & {\color{gray}96.23}   & {\color{gray}10.09 $\pm$ 10.02} & {\color{gray}0.56 $\pm$ 1.16}   \\
 & {\color{purple}GGFast}       & {\color{purple}74.31}     & {\color{purple}79.58}     & {\color{purple}50} & {\color{purple}1.82 $\pm$ 0.92} \\
& {\color{purple}Grad-BP} & {\color{purple}90.62} & {\color{purple}93.48} & {\color{purple}100(TCP) 10(UDP)} & {\color{purple}3.71 $\pm$ 1.33} \\
                             & Pkt.-5 & 80.31 & 88.85 & 5 $\pm$ 0.00 & 0.32 $\pm$ 0.96  \\
                             & Pkt.-45         &89.55     &95.7     &45 $\pm$ 0.00 & 1.61 $\pm$ 0.81 \\
                             & Time-int.-5 & 69.76 & 73.80 & 2.12 $\pm$ 2.42 &.25  \\
                             & Time-int.-45 & 91.29 & 92.94 & 63.96 $\pm$ 7.97 & 2.25 \\

\hline

\end{tabular}
\end{table}
Table~\ref{tab:results_table_packet_drop} shows the evaluation results when the datasets are augmented with packet sequence disorders. 
It can be seen that \textit{FastFlow} achieves a balanced performance in both classification accuracy and speed compared to all methods. Classifiers that only use packet data (\ie `Packet seq', `Pkt.-5' and `Pkt.-45') are not robust to packet sequence disorders. Classifiers that only use time-interval slot data (\ie `Time-int seq.', `Time-int.-5' and `Time-int.-45') are either not having good accuracy with small numbers of inputs or requiring long time (\eg over 45 seconds) for flow classification.

\subsubsection{With Unknown Flow Types but not Packet Sequence Disorder}
\begin{table}[t!]
\centering
\caption{Classification performance of \textit{FastFlow} with unknown flow type.}
\label{tab:results_table_ood}
\fontsize{7}{9}\selectfont
\begin{tabular}{|l|l|c|c|c|c|c|c|}
\hline
\multirow{2}{*}{\textbf{Dataset}} & \multirow{2}{*}{\textbf{Method}}  & \multicolumn{4}{c|}{\cellcolor[rgb]{ .906,  .902,  .902}\textbf{Classification Performance}} & \multicolumn{2}{c|}{\cellcolor[rgb]{ .906,  .902,  .902}\textbf{Unk. flow detection}} \\ \cline{3-8}

                                  &                                   & \textbf{Macro F1 (\%)} & \textbf{Accuracy (\%)} & \textbf{Packets (\#)} & \textbf{Time (s)} & \textbf{FPR (\%)} & \textbf{TPR (\%)} \\ \hline
\multirow{9}{*}{UTMobileNet}       

& \textbf{\color{blue}FastFlow}                         & {\color{blue}89.21}   & {\color{blue}90.2}      & {\color{blue}12.74 $\pm$ 5.99} & {\color{blue}0.59 $\pm$ 1.07} & {\color{blue}4.16}          & {\color{blue}88.21}            \\ 
& {\color{gray}Packet seq.}                                & {\color{gray}90.07}             & {\color{gray}90.7}              & {\color{gray}8.51 $\pm$ 3.63} & {\color{gray}0.41 $\pm$ 0.23} & {\color{gray}2.81}          & {\color{gray}90.01}            \\
                                 
 &{\color{gray}Time-int. seq.}                                 & {\color{gray}89.33}             & {\color{gray}89.72}  & {\color{gray}29.63 $\pm$ 10.29} & {\color{gray}1.68 $\pm$ 3.79} & {\color{gray}4.59}         & {\color{gray}85.38}            \\ 

   & {\color{purple}GGFast} & {\color{purple}90.35} & {\color{purple}91.49} & {\color{purple}50} & {\color{purple}2.92 $\pm$ .96} & {\color{purple}4.97} & {\color{purple}92.81} \\
 & {\color{purple}Grad-BP} & {\color{purple}91.21} & {\color{purple}91.68} & {\color{purple}100(TCP) 10(UDP)} & {\color{purple}5.73 $\pm$ 1.49} & {\color{purple}4.88} & {\color{purple}69.87} \\

    &  Pkt.-5 & 72.95 & 77.78  & 5  & 0.15 $\pm$ 1.99 & -- & --  \\
    & Pkt.-45 &  88.96  & 90.07 & 45 & 2.61 $\pm$ 0.79 &-- & -- \\
    & Time-int.-5 & 69.51 & 74.93 & 6.49 $\pm$ 2.94 & 0.25& -- & -- \\
    & Time-int.-45 & 88.85 & 89.49  & 39 $\pm$ 6.04 & 2.25 & -- & -- \\

\hline
\multirow{9}{*}{VNAT}             
 & \textbf{\color{blue}FastFlow}   & {\color{blue}99.33}  & {\color{blue}99.46}  & {\color{blue}3.89 $\pm$ 1.61} & {\color{blue}.06 $\pm$0.74}  & {\color{blue}0}            & {\color{blue}99.39}            \\ 
    & {\color{gray}Packet seq.}                               & {\color{gray}99.33}             & {\color{gray}99.46}             & {\color{gray}3.89 $\pm$ 1.61} & {\color{gray}.06 $\pm$0.07} & {\color{gray}0}  & {\color{gray}99.20}              \\
    & {\color{gray}Time-int. seq.}                                & {\color{gray}99.70}             & {\color{gray}99.80}   & {\color{gray}7.85 $\pm$ 4.35} & {\color{gray}.088 $\pm$ 1.64} & {\color{gray}0}  & {\color{gray}98.93}              \\
   & {\color{purple}GGFast} & {\color{purple}84.28} & {\color{purple}90.72} & {\color{purple}50} & {\color{purple}0.95 $\pm$ .62} & {\color{purple}2.79} & {\color{purple}99.42} \\
    & {\color{purple}Grad-BP} & {\color{purple}99.77} & {\color{purple}99.79} & {\color{purple}100(TCP) 10(UDP)}& {\color{purple}1.97 $\pm$ 0.84} & {\color{purple}4.98} & {\color{purple}99.30} \\

&  Pkt.-5 & 92.01 & 96.50 & 5  & 0.03 $\pm$ 0.83 & -- & --  \\
    & Pkt.-45 &  94.72  & 98.67 & 45 & 0.84 $\pm$ 0.59 &-- & -- \\
    & Time-int.-5 & 93.42 & 98.96 & 15.79 $\pm$ 3.07 & 0.25& -- & -- \\
    & Time-int.-45 & 90.43 & 93.50  & 93.52 $\pm$ 11.96 & 2.25 & -- & -- \\

\hline
\multirow{9}{*}{UNIBS}      
 & \textbf{\color{blue}FastFlow}                         & {\color{blue}94.56}             & {\color{blue}97.80}             & {\color{blue}6.32 $\pm$ 2.68} & {\color{blue}0.28 $\pm$ 1.30} & {\color{blue}1.33}          & {\color{blue}97.32}            \\ 
  & {\color{gray}Packet seq.}                               & {\color{gray}95.10}             & {\color{gray}97.79}             & {\color{gray}5.19 $\pm$ 3.02} & {\color{gray}0.23 $\pm$ 1.16}  & {\color{gray}0}             & {\color{gray}98.29}            \\
 & {\color{gray}Time-int. seq.}                                 & {\color{gray}92.62}             & {\color{gray}96.18}             & {\color{gray}11.41 $\pm$ 7.81} & {\color{gray}0.39 $\pm$ 0.83} & {\color{gray}4.18}          & {\color{gray}98.40}            \\ 

& {\color{purple}GGFast} & {\color{purple}93.12} & {\color{purple} 97.45}& {\color{purple}50} & {\color{purple}1.35 $\pm$ 1.31} & {\color{purple}3.43} & {\color{purple}99.03} \\
& {\color{purple}Grad-BP} & {\color{purple}89.09} & {\color{purple}96.11} & {\color{purple}100(TCP) 10(UDP)} & {\color{purple}2.53 $\pm$ 1.63} & {\color{purple}4.90} & {\color{purple}73.90} \\

    & Pkt.-5 & 85.08  &  89.81 & 5  & 0.23 $\pm$ 0.93 & -- & --  \\
    & Pkt.-45 &  93.56  & 97.54 & 45 &1.22 $\pm$ 1.84 &-- & -- \\
    & Time-int.-5 & 82.45 &  87.01 & 5.62 $\pm$ 1.92 & 0.25& -- & -- \\
    & Time-int.-45 & 92.73 & 95.63  & 86.01 $\pm$ 21.02 & 2.25 & -- & -- \\
                                
\hline
\end{tabular}
\end{table}

Table~\ref{tab:results_table_ood} reports the evaluation results on the three datasets augmented with unknown flows. Compared to the two state-of-the-arts methods and \textit{FastFlow} classifiers with only packet or time-interval slot data sequences, \textit{FastFlow} achieves decent performance in both classification accuracy and speed for both known and unknown flows.

\end{document}